\documentclass[12pt,preprint,twocolumns]{emulateapj}
\usepackage{latexsym}
\usepackage{url}
\usepackage{color,ulem}
\usepackage[hypertex]{hyperref}
\usepackage{amssymb}
\usepackage{amsmath}
\usepackage{epsfig}
\usepackage{appendix}
\usepackage{wrapfig}
\singlespace

\def\CII{\hbox{\rm C~$\scriptstyle\rm II$}}
\def\OI{\hbox{\rm O~$\scriptstyle\rm I$}}

\def\kms{\,{\rm km\,s^{-1}}}

\def\cc{\,{\rm cm^{-3}}}

\def\msun{\,{\rm M_\odot}}

\def\Lya{Ly$\alpha$}

\def\spose#1{\hbox to 0pt{#1\hss}}
\def\lta{\mathrel{\spose{\lower 3pt\hbox{$\mathchar"218$}}
     \raise 2.0pt\hbox{$\mathchar"13C$}}}
\def\gta{\mathrel{\spose{\lower 3pt\hbox{$\mathchar"218$}}
     \raise 2.0pt\hbox{$\mathchar"13E$}}}
\def\ni{\noindent}

\lefthead{Madau et al.}
\righthead{Formation of GCs in Subhalo-Subhalo Collisions}
\submitted{}


\begin{document}

\title{Globular Cluster Formation from Colliding Substructure}

\author{Piero Madau\altaffilmark{1}, Alessandro Lupi\altaffilmark{2}, J{\"u}rg Diemand\altaffilmark{3}, Andreas Burkert\altaffilmark{4,5}, 
and Douglas N. C. Lin\altaffilmark{1}}

\affil{$^1\,$Department of Astronomy \& Astrophysics, University of California, 1156 High Street, Santa Cruz, CA 95064, USA\\
$^2\,$Scuola Normale Superiore, Piazza dei Cavalieri 7, I-56126 Pisa, Italy\\
$^3\,$Cisco Systems (Switzerland) GmbH, Richtistrasse 7, 8304 Wallisellen, Z{\"u}rich, Switzerland\\ 
$^4\,$ University Observatory Munich (USM), Scheinerstrasse 1, D-81679 Munich, Germany\\ 
$^5\,$ Max-Planck-Institut f\"ur Extraterrestrische Physik (MPE), Giessenbachstr. 1, D-85748 Garching, Germany 
}

\begin{abstract}
We investigate a scenario where the formation of Globular Clusters (GCs) is triggered  by high-speed collisions between 
infalling atomic-cooling subhalos during the assembly of the main galaxy host, a special dynamical mode of star formation that operates at high 
gas pressures and is intimately tied to $\Lambda$CDM hierarchical galaxy assembly. The proposed mechanism would give origin to ``naked" globulars,
as colliding dark matter subhalos and their stars will simply pass through one another while the warm gas within them clashes at highly supersonic 
speed and decouples from the collisionless component, in a process reminiscent of the Bullet galaxy cluster. We find that the resulting shock-compressed 
layer cools on a timescale that is typically shorter than the crossing 
time, first by atomic line emission and then via fine-structure metal-line emission, and is subject to gravitational instability and fragmentation.
%We use the thin-shell approximation to investigate the requirements for imprinting the signature of the GC mass scale on the collapsing shell.
Through a combination of kinetic theory approximation and high-resolution $N$-body simulations, we show that this model may produce: 
{\it (a)} a GC number-halo mass relation that is linear down to dwarf galaxy scales and agrees with the trend observed over five orders of magnitude in galaxy mass; 
{\it (b)} a population of old globulars with a median age of 12 Gyr and an age spread similar to that observed; {\it (c)} 
a spatial distribution that is biased relative to the overall mass profile of the host; 
%This is because, in an inelastic collision, the splash remnant will lose orbital energy and fall deeper into the Galactic potential rather than sharing the 
%orbits of the progenitor subhalos;
and {\it (d)} a bimodal metallicity distribution with a spread similar to that observed in massive galaxies. 
%Most interacting subhalos have correlated infall histories, highly radial orbits, and plunge deep into the main host.
%Additional, hydrodynamic simulations of subhalo-subhalo violent impacts should be performed to further validate a collision-driven scenario for the formation of GCs.
\end{abstract}


\section{Introduction}

The origin of globular clusters (GCs) remains an unsolved problem in star and galaxy formation studies. With masses in the range 
$10^4$--$10^6\,\msun$, half-light radii of a few pc, a bimodal metallicity distribution, internal (star-to-star) variation in 
their light-element abundances, and typical ages $>$10 Gyr, these remarkably compact stellar systems are a common feature of 
galaxies in the local Universe \citep[see, e.g., the reviews by][]{harris01,west+04,brodie+06,gratton+12,kruijssen14,forbes18}. 
Young massive stellar clusters, with masses and densities comparable to those of GCs, are observed in galaxy mergers throughout the local 
Universe \citep[e.g.,][]{whitmore+99}, suggesting that the progenitors of metal-rich globulars could still be forming today in unusually 
high pressure environments. The mass of the entire GC system of a galaxy and its radial extent correlate tightly 
with the dark matter (DM) mass of the host halo \citep[see, e.g.,][and references therein]{harris15,forbes16,harris17,hudson18}, 
a trend the former suggestive of a picture in which star formation in dense proto-GCs was relatively immune to the feedback mechanisms that hamper most star formation in the field.

The Milky Way 
(MW) contains about 160 known GCs. The blue (metal-poor) halo population is more extended, with a spatial distribution that falls off 
as $r^{-3.6\pm 0.2}$ at Galactocentric distances $\gta 3.5$ kpc \citep{bica+06}. Only six known GCs inhabit the Galaxy at distances $\gta 
90$ kpc \citep[][]{harris96,laevens14}. The red (metal-rich) globulars are more spatially concentrated than the blue clusters: they lie 
within the solar circle and form a flattened, rotating population. Blue GCs outnumber red globulars by a ratio of 3:1. Despite their 
different location and kinematics, the red and blue clusters appear to have similar internal properties, such as masses, sizes, and ages. 
The age spread among the bulk of globulars is about $2$ Gyr, but there are a few outliers \citep[e.g.,][]{leaman13}. The youngest GC, 
Whiting 1, is metal-rich and has an estimated age of 
$5.7\pm 0.3$ Gyr \citep{valcheva15}, similar to those of Pal 1 and Terzan 7. The newly discovered metal-poor GC in Crater 
\citep{belokurov14,laevens14}, located at a distance of $145\pm 0.3$ kpc, is well-described by a simple stellar population with an age of 
$7.5\pm 0.4$ Gyr \citep{weisz16}. While most Galactic GCs formed before the peak of the cosmic star-formation rate density at redshift 2 
\citep{maddick14}, a handful were clearly assembled as late as $z\lta 1$. 

It has been an enduring challenge to explain such features, and despite a wide variety of proposed phenomenological models and a flurry of 
work on the present-day properties and evolution of GCs in galaxy halos, a self-consistent scenario of GC formation is yet to be constructed.
It has long been suggested, for example, that a significant fraction of MW's GCs formed {\it ex-situ} during the very early stages of 
galaxy formation, typically in dwarf-sized systems that were subsequently accreted into the Galactic potential well 
\citep[][but see \citealt{chiou19}]{searle+78,peebles84,rosenblatt+88,cen01,bromm+02,diemand05,moore06,boley09,griffen+10,trenti+15,kimm16,ricotti16,boylan17,creasey19}.
By often relying on ad-hoc prescriptions for the formation of globulars inside their own DM halos, these ``pregalactic" models have been 
somewhat successful in reproducing some puzzling properties of the GC populations. The observed spatial coincidence between GCs 
and multiple tidal debris streams in the outer halo of M31 does indeed suggest a direct connection between some GCs and dwarf 
galaxy remnants \citep{mackey10a}.  
%A pregalactic, $z\gta 5$, origin for the blue GC population would predict, however, an age spread $\lta 1$ Gyr. 
{\it In-situ} mechanisms have focused instead on the formation of GCs inside the main progenitor of their present-day host,
perhaps from cooling-induced fragmentation of low-metallicity infalling gas \citep{fall85}, during the merger with a gas-rich massive companion
\citep{ashman92,muratov10,choksi18}, or within the dense cores of giant molecular clouds in early galactic disks \citep{kravtsov+05,kruijssen15,pfeffer18}. 

Most clusters in the MW contain no measurable amounts of DM. The faint tidal tails observed around some GCs 
\citep{grillmair+95,odenkirchen+03} provide strong constraints on their mass-to-light ratios and 
indicate that their mass distribution does not extend beyond the optical radius \citep{moore96}. The absence of DM 
could be the result of stripping in the strong tidal field of the Galaxy 
\citep{mashchenko+05,saitoh+06}, and does not preclude by itself the formation of GCs at the center of DM substructure. {\it Ex-situ} 
scenarios in which GCs form at the center of their DM halos do predict, however, 
detectable amounts of DM in the outskirts of GCs that have always resided within the weak tidal field of the outer MW halo. This is 
not borne out by the data: recent observations of stellar kinematics in NGC 2419, located at 90 kpc from the center of our Galaxy, 
and in MGC1, located at 200 kpc from the center of M31, show that these GCs cannot be deeply embedded within dark halos having a 
virial mass greater than $10^6\,\msun$ \citep{baumgardt+09,conroy+11,ibata13}. 

In this paper we propose and investigate a possible scenario for the formation of GCs within the framework of $\Lambda$CDM hierarchical 
galaxy assembly. Many GC formation models leave largely unanswered the fundamental question of how, at early cosmological times, 
self-bound and extremely dense aggregates of $\sim 10^5$ stars, largely mono-metallic in iron-peak elements, were able to form 
and survive without retaining much of the DM of their former host. GCs are the result of a special mode of star formation 
that requires extremely high gas pressures, $p\sim 10^7\,k_B$ cm$^{-3}\,$K, some 3 dex higher than typical interstellar values. 
It is these pressures 
that hinder the dispersal of star-forming material, protect dense proto-GCs from the feedback processes that regulate star 
formation in the field, and produce high star formation efficiencies \citep[e.g.,][]{elmegreen+97}. They are the kind of 
pressure that would result, e.g., if typical atomic clouds in the interstellar medium (ISM) with densities of a few atoms cm$^{-3}$ 
were to collide at the Galactic orbital speeds of $200\,\kms$, which is precisely the outcome of the mechanism proposed here: 
{\it early high-speed collisions between subhalos infalling onto massive galaxy hosts, leading to the formation of 
dark-matter-less GCs}. 

\section{Subhalo-Subhalo Collisions}

Numerical simulations in $\Lambda$CDM have shown that massive galaxy halos are populated by a rich spectrum of substructures that collapsed at 
early times, the fossil remnants of a hierarchical merging process that is never complete 
\citep{moore+99,klypin+99,diemand+07,diemand08,springel08}. About one thousand subhalos with pre-infall masses $m_{\rm peak}\gta 10^{8}\,\msun$ 
are predicted to have been accreted by our Galaxy over cosmic 
time \citep[see][and references therein]{han18}.\footnote{To be more precise, $m_{\rm peak}$ is the maximum mass attained by a subhalo 
over its entire merger history. This mass is greater than the subhalo mass at infall, as most subhalos start being stripped 
at $\lta 2$ host virial radii, regardless of host mass \citep{behroozi14}. Infalling halos generally become accreted shortly 
after reaching peak mass.}\, While cold gas condensation and normal, Population II star formation 
can take place at high redshift in these ``atomic-cooling" subhalos because of hydrogen atomic radiative processes, 
their overall efficiency of converting baryons into stars must remain low in order to avoid overproducing the observed abundance of dwarf 
galaxy satellites of the MW \citep{madau08,boylan-kolchin14}. Mechanisms such as early reionization of the intergalactic medium, supernova 
feedback, and H$_2$-regulation have all been invoked in order to reduce the star formation efficiencies of low-mass field halos and even 
prevent the smallest ones from forming stars altogether, and ``dark dwarfs" with long gas depletion timescales are often produced in cosmological 
simulations \citep[e.g.,][]{okamoto08,kuhlen13,shen14,benitez15,sawala15,ricotti16}.  Reionization heating is expected to photo-evaporate 
low-density gas from these shallow potential wells, but to have little effect on relatively high-density gas in their cores 
that can self-shield from UV background radiation. The detailed mapping betwen the mass spectrum of DM substructure and the population of 
Galactic satellites remains, however, a matter of debate, and so is the smallest mass halo that is capable of hosting an observable baryonic 
counterpart. A recent probabilistic comparison (abundance matching) between the luminosity function of MW's dwarfs and $N$-body zoom-in 
simulations sets a $1\sigma$ upper bound to the pre-infall peak subhalo mass of Segue I -- the faintest MW satellite -- of $m_{\rm peak}<
2.4\times 10^8\,\msun$ \citep{jethwa18}. Similarly, a fresh reassessment of the ``missing satellites problem", cast in terms of number counts, 
suggests that luminous satellites of the MW inhabit subhalos with infall masses as small as $10^7-10^8\,\msun$ \citep{kim18}. 

The central idea of this work is to suggest that high-speed collisions between infalling substructures during the assembly of the main
galaxy host may result in the formation of ``naked" GCs. Like in the Bullet galaxy gluster \citep{clowe06}, colliding DM subhalos
and their stars will simply pass through one another, largely continuing on their original orbital path (close encounters are more
effective at disrupting the colliding satellites at low speeds). The atomic gas within them, however, will collide at highly
supersonic speed and decouple from the collisionless component. The collision will compress and shock-heat the gas
to characteristic temperatures $\gta 10^5\,$K, creating the very high pressure region that is conducive to GC formation.
We will show that the low-metallicity gas will cool via atomic line emission on a timescale that is much shorter
than the crossing time. Under the appropriate conditions, shock dissipation of the relative kinetic energy will lead to the
coalescence of the colliding gas clouds and the formation of a Jeans unstable slab, rather than cloud disruption. A high star
formation efficiency may result from the short dynamical timescale and high binding energy of the splash remnant.

\subsection{Kinetic Theory Approach}

An estimate of the number of close collisions can be made within the framework of kinetic theory \citep[e.g.,][]{makino97}. 
Let us start by selecting subhalos by their pre-infall mass, i.e. by using the ``unevolved" substructure mass function -- the 
distribution of peak bound masses $m_{\rm peak}$ of subhalos accreted by the host at all previous redshifts \citep[e.g.,][]{vdB05,jiang16}.
In $\Lambda$CDM cosmological simulations, this universal redshift-independent function is well fitted by a double Schechter 
function of the form \citep{han18}
\begin{equation}
{dN_{\rm sub}\over d\xi}=\xi^{-1}\left(a_1\xi^{-\alpha_1}+a_2\xi^{-\alpha_2}\right)\exp(-b\xi^{\beta}),
\label{eq:dNdmu}
\end{equation}
where $\xi=m_{\rm peak}/M_{\rm vir}$ is the ratio of the peak mass of a subhalo to the host halo mass, 
and $(dN_{\rm sub}/d\xi)d\xi$ is the number of subhalos with masses between $\xi$ and $\xi+d\xi$.  When adopting for the 
host mass the virial definition corresponding to the overdensity of the spherical collapse model, the best-fit parameters are $a_1=0.11, 
a_2=0.32, \alpha_1=0.95, \alpha_2=0.08, b=8.9,$ and $\beta=1.9$ \citep{han18}. In the case of small subhalos, 
the unevolved spatial distribution (the spatial distribution at fixed infall mass, see \citealt{han16}) follows
the spherical \citet{navarro97} (NFW) density profile of the host,
\begin{equation}
n_{\rm sub}(\xi,x)\propto {1\over cx(1+cx)^2}, 
\label{eq:dNdV}
\end{equation}
where $x\equiv r/R_{\rm vir}$, $R_{\rm vir}$ is the virial radius of the host, and $c$ is 
its ``concentration parameter".\footnote{In reality, the dynamics of subhalos differs somewhat from that of DM particles because of dynamical      
friction, but this is a small effect for the majority of subhalos that are located in the outer regions of the main host and at the low-mass end \citep{han16}.}\,
The subhalo unevolved mass function and spatial distributions are approximately 
separable (i.e., the spatial distributions of different peak mass subhalos are similar except for a change in amplitude), and 
the number density of small subhalos with peak masses $>\xi M_{\rm vir}$, $n_{\rm sub}(>\xi,x)$, can be written as
\begin{equation}
4\pi R_{\rm vir}^3 n_{\rm sub}(>\xi,x)= {c^3\over gcx(1+cx)^2}\int_{\xi}^1 {dN_{\rm sub}\over d\xi'} d\xi', 
\label{eq:nsub}
\end{equation}
where $g\equiv 1/[\ln(1+c)-c/(1+c)]$. 

Let us further make the simplifying assumption that the velocity distribution of subhalos is a local Maxwellian with 
one-dimensional velocity dispersion $\sigma_v$. The distribution of relative encounter speeds is then 
\begin{equation}
f(v_{\rm rel})={1\over 2\sqrt{\pi}\sigma_v^3}v_{\rm rel}^2\exp\left(-{v_{\rm rel}^2\over 4\sigma_v^2}\right),
\end{equation}
and the mean encounter speed is $\bar v_{\rm rel}=4\sigma_v/\sqrt{\pi}$. For isotropic orbits in an NFW potential,
the velocity dispersion obtained by solving the Jeans equation is \citep{lokas01}
\begin{equation}
\begin{split}
\sigma_v^2(x)= & \left({GM_{\rm vir}\over R_{\rm vir}}\right)\,g(1+cx)^2x \\
& \times \int_x^\infty \left[{\ln(1+cs)\over s^3(1+cs)^2}-{c\over s^2(1+cs)^3}\right]ds.
\end{split}
\label{eq:sigmav}
\end{equation}
Under the assumption that collisions occur on random orbits,  
we can finally estimate the mean number of encounters between subhalo pairs of peak masses (in units of $M_{\rm vir}$) 
$>{\xi}_1$ and $>{\xi}_2$ in a time interval $\Delta t$ as
\begin{equation}
\begin{split}
N_{\rm coll}= & \Delta t\,4\pi R_{\rm vir}^3\int_0^1 \bar v_{\rm rel}(x)x^2 dx \\ 
& \times \int_{{\xi}_1}^1 \int_{{\xi}_2}^1 n_{\rm sub}(\xi,x) n_{\rm sub}({\xi}',x)(\pi b_{\rm max}^2)d\xi d{\xi}', 
\end{split}
\label{eq:dotN}
\end{equation}
where $b_{\rm max}$ is the maximum impact parameter for a  close collision. 

Consider now, for illustrative purposes, a MW-sized host halo of mass $M_{\rm vir}=10^{12}\,\msun$ and 
concentration $c=10$ \citep[e.g.,][]{duffy08}. Its isotropic velocity dispersion profile in Equation 
(\ref{eq:sigmav}) reaches a maximum value of $102\,\kms$ at $x=0.08$, 
corresponding to a mean collision speed of $230\,\kms$. These relative orbital velocities 
are much greater than the internal velocities of subhalos, and few such encounters will actually result 
in mergers. We can therefore neglect gravitational focusing and, in order to ensure that the region of 
overlap between the gaseous cores of the interacting clumps is extensive, define penetrating collisions as those 
with $b_{\rm max}=0.1(r_{\rm vir,1}+r_{\rm vir,2})$, 
where $r_{{\rm vir},1}$ and $r_{\rm vir,2}$ are the pre-infall virial radii of the two clumps. 
The mean number of impacts within $R_{\rm vir}$ between subhalos having peak masses 
$m_{\rm peak}>10^8\,\msun$ is then
\begin{equation}
N_{\rm coll}\simeq 59\,\left({\Delta t\over {\rm Gyr}}\right). 
\label{eq:dotNis}
\end{equation}
This is clearly an interesting figure -- comparable after $\Delta t\sim$ a few Gyr to the number of GCs observed 
today in the halo of the MW -- if many such encounters were to result in the formation of one or more globulars.
Because of the steep substructure mass function, most collisions involve subhalos at the small-mass end 
of the distribution: the mean frequency of close encounters between atomic-cooling subhalos 
with $m_{\rm peak}>10^8\,\msun$ 
and satellites that are ten times more massive is 3.6 times smaller than Equation (\ref{eq:dotNis}). 
There are only $N_{\rm coll} \sim 3\,(\Delta t/{\rm Gyr})$ close collisions between massive 
satellites with $m_{\rm peak}>10^9\,\msun$, but these may give origin to multiple GCs per encounter. The 
same calculation predicts about 1 collision Gyr$^{-1}$ between atomic-cooling subhalos in a dwarf galaxy-sized host with $M_{\rm vir}=10^{10}\,\msun$.
 
In reality, of course, many of the simplified assumptions behind our integration of Equation (\ref{eq:dotN}) do not hold in
practice: 1) subhalos are often accreted at similar times and locations as members of groups along filaments, and this
causes an enhancement of encounters at small angles \citep{benson05,zentner05,gonzalez16}; 2) after accretion, subhalos
are susceptible to dynamical friction and tidal stripping, and their mass and spatial distributions 
evolve away from their form at infall \citep{diemand07,springel08}; 3) the abundance of subhalos above a given peak mass 
evolves with time, and so does the background gravitational potential in which they move; 4) subhalos are predicted to have 
a non-Maxwellian orbital velocity function, with centrally rising velocity anisotropy \citep{sawala17}; 
and 5) ram-pressure stripping will remove the outer gaseous component of subhalos on orbits with small
pericenter radii, a process that is most effective close to pericenter passage \citep{mayer06,grcevich09}.
In \S\,\ref{S:VLI} we shall tackle some of these issues by using the densely time-sampled snapshots of the {\it Via Lactea}
$N$-body simulation to track collisions between infalling substructure in a massive MW-sized host that grows and evolves in
a fully cosmological context. 

%Subhalos on radial orbits that pass close to the host center may also be depleted 
%by ``disk shocking" \citep{donghia10,garrison-kimmel17}. 

%
\begin{figure}
\plotone{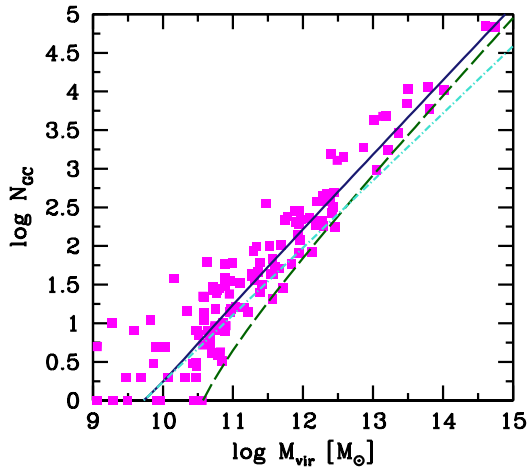}
\caption{The GC system number-host halo mass relation. The data points are observational estimates from a recent compilation by 
\citet{burkert19}. The lines show the close-to-linear relation predicted by a kinetic theory model of cluster formation following
subhalo-subhalo collisions (the normalization is arbitrary, see text for details).
Solid curve: Mean number of GCs formed either in the main progenitor of a halo of mass $M_{\rm vir}$ ({\it in-situ} globulars) 
or in a companion system and later accreted ({\it ex-situ} globulars).
{Dashed curve:} {\it Ex-situ} globulars only. 
{Dot-dashed curve:} {\it In-situ} globulars only. 
} 
\label{figNgc}
\vspace{+0.5cm}
\end{figure}

\subsection{Scaling with Host Mass and Ex-situ Globulars}

One of the most intriguing aspects of GC phenomenology is their relationship to DM halos: the total
mass of the GC population of a galaxy and the number of GCs correlate linearly with the DM mass of the host halo.
This trend is valid over five orders of magnitude in galaxy mass \citep[see, e.g.,][and references
therein]{harris15,forbes16,burkert19}, and contrasts markedly with the non-linear relation between total stellar mass and
DM-halo mass \citep[e.g.,][]{behroozi13}. The observed correlation may point to a fundamental GC system-DM connection 
rooted in the cluster formation physics, or simply be the inevitable consequence of hierarchical assembly and the 
central limit theorem \citep{boylan17,el-badry19,burkert19}.

It is of obvious interest at this stage to use kinetic theory and derive a 
scaling relation between the number of collisions and the virial mass of the host following Equation (\ref{eq:dotN}). 
The collision frequency between subhalos more massive than a given $m_{\rm 
peak}\ll M_{\rm vir}$ at fixed concentration parameter is proportional to $n_{\rm sub}^2\propto R_{\rm vir}^{-6}\,M_{\rm vir}^{1.9}$ 
(from Eqs. \ref{eq:dNdmu} and \ref{eq:nsub})
times $\bar v_{\rm rel}\propto (M_{\rm vir}/R_{\rm vir})^{1/2}$ (see Eq. \ref{eq:sigmav}). Including a dependence on halo 
concentration and assuming that the collisions of interest continue for a timescale $\Delta t$ that is independent of host mass, 
one infers from Equation (\ref{eq:dotN}):
\begin{equation}
N_{\rm coll} \propto R_{\rm vir}^{-3}\,M_{\rm vir}^{1.9}\,(M_{\rm vir}/R_{\rm vir})^{1/2} c^4.
\label{eq:Ncoll}
\end{equation}
Noting that $R_{\rm vir}\propto M_{\rm vir}^{1/3}$ and using the halo concentration-mass relation of \citet{duffy08},
%(relaxed $z=$0-2 halo sample)
$c\propto M_{\rm vir}^{-0.09}$, we obtain the following scaling $N_{\rm coll} \propto M_{\rm vir}^{0.87}$.
Under the assumption that GCs populate present-day DM halos in direct proportion to $N_{\rm coll}$, 
let us write  
\begin{equation}
N^{\rm GC}_{\rm in-s}=A\,(M_{\rm vir}/10^{12}\,\msun)^{0.87},
\label{eq:Nins}
\end{equation}
where $A$ is an arbitrary normalization factor. In this scenario, $N^{\rm GC}_{\rm in-s}$ tracks the population of 
GCs formed {\it in-situ} within a massive galaxy host. 
The same substructure collision mechanism, at work in satellites prior to infall, would also produce a 
significant population of globulars that formed {\it ex-situ} and were subsequently accreted.
The $N^{\rm GC}_{\rm in-s}-M_{\rm vir}$ relation given above and the substructure mass function in Equation (\ref{eq:dNdmu})
allow a straightforward calculation of the mean number of accreted GCs as
\begin{equation}
N^{\rm GC}_{\rm ex-s}=N^{\rm GC}_{\rm in-s}\,\int_{\xi_{\rm min}}^1 \xi^{0.87} {dN_{\rm sub}\over d\xi} d\xi,
\label{eq:Nacc}
\end{equation}
where $\xi_{\rm min}\equiv m_{\rm min}/M_{\rm vir}$ is the critical subhalo mass below which GCs cannot form. Obviously, 
collisions between $m_{\rm peak}>10^8\,\msun$ clumps as a mechanism for cluster formation require
at least a few (say $\sim 4$) atomic-cooling subhalos in a given host; this trivial condition yields, after integrating Equation (\ref{eq:dNdmu}), 
$m_{\rm min}\simeq 10^{9.6}\,\msun$. This critical mass could be larger -- thereby decreasing the fraction 
of accreted globulars -- if the relative orbital velocity $v_{\rm rel}$ required to trigger the formation of a GC 
by impact had to exceed $\sim 50\,\kms$. Note, however, that lower velocity encounters, as those expected in a less massive host, 
may still produce the high pressures that are conducive to GC formation if the colliding gaseous cores were 
typically denser, e.g. at very high redshifts. 

Assuming here that all halos below $m_{\rm min}=10^{9.6}\,\msun$ do not contain a GC,
Equation (\ref{eq:Nacc}) gives $N^{\rm GC}_{\rm ex-s}/N^{\rm GC}_{\rm in-s}=0.35,0.73,1.17$ for $M_{\rm vir}=10^{11},10^{12},10^{13}\,\msun$,
i.e. {\it most GC formation occurs {\it in-situ} for the lowest-mass host halos and most GCs in massive elliptical systems are 
actually accreted.}\footnote{Similar trends are seen in recent phenomenological models of GC formation based on DM merger trees 
\citep[e.g.][]{boylan17}.}\,  
With the addition of the accreted globulars, the scaling of the {\it total} number of GCs, $N_{\rm GC}=N^{\rm GC}_{\rm ex-s}+
N^{\rm GC}_{\rm in-s}$, with host virial mass becomes very close to linear, 
\begin{equation}
N_{\rm GC}\propto M_{\rm vir}^{0.98}.
\label{eq:NGC}
\end{equation}
{\it Therefore, in a collision-driven scenario, a constant GC number-to-halo mass ratio is the result
of encounter probability calculations (Eq. \ref{eq:Ncoll}) and hierarchical clustering (Eq. \ref{eq:Nacc}): 
GCs are the result of a distinctive mode of star formation that tracks the total mass of their host galaxy 
rather than its stellar mass.} 

Figure \ref{figNgc} shows the observed GC number-halo mass relation from a recent compilation by \citet{burkert19} 
(see also \citealt{forbes18dg,harris17}), compared with our theoretical prediction for an assumed 
normalization $A=95$ in Equation (\ref{eq:Nins}).
The model appears to reproduce the observed trend reasonably well, with the expected average
number of GC systems dropping below a few in dwarf galaxies with $M_{\rm vir}\lta 10^{10}\,\msun$. 
With the adopted parameters, a MW-mass system with $M_{\rm vir}=10^{12}\,\msun$ today 
would host a grand total of 165 massive GCs, of which 70 formed {\it ex-situ}. Estimates of the number of accreted GCs within the MW today 
range from 30 to 90 \citep{forbes10,leaman13}, so it is conceivable that a substantial fraction of globulars in massive hosts 
may have an external origin. The significant scatter in the observed relation on dwarf galaxy scales may reflect 
variations in merger histories, uncertainties in dark halo mass determinations based on kinematical tracers, 
or environmental effects such as tidal stripping \citep{burkert19}.

While it is reassuring that a scenario in which GCs are the result of collisions between subhalos may
be able to accomodate a uniform GC production rate per unit host halo mass as implied by the data, 
it may be premature to read too much into this comparison given the above-mentioned limitations of kinetic theory. 
In particular, our modeling so far has not provided any information on the age and age-spread of GCs, on the epoch at which 
the GC-to-halo mass relation may actually be established, on the normalization of such relation, and on the 
baryonic physics leading to the formation of massive globulars. In the following sections, we shall discuss the conditions 
under which high-speed impacts may lead to cloud coalescence, gravitational instability, and the formation of 
``naked", bound GCs, and use numerical simulations to argue that a collision-based framework may fulfill several key 
observational constraints on GCs that have emerged over the last two decades.

\section{GC Formation in High-Speed Impacts}

Observations of interacting galaxies like the Antennae and Mice pairs show widespread stellar cluster formation
generated by the collision, and high-resolution simulations of encounters between disk galaxies with realistic ISM reveal
the presence of shock-induced gas compression and star formation at the collision interface \citep[e.g.,][]{saitoh09}.
There is a vast literature on supersonic cloud-cloud collisions as a possible triggering mechanism for star formation in the ISM
\citep[e.g.,][]{stone70,smith80,gilden84,nagasawa87,habe92,anathpindika09,arreaga14,balfour17}, and cloud-cloud collisions in the young
Galaxy have already been invoked in the context of GC formation \citep[e.g.,][]{murray89,kang90,lin91,kumai93,hartwick09}. Below, we show that, when a similar collision occurs
between subhalos, the post-shock gas remnant will cool down on a timescale that is much shorter than the compression time,
and may, under the right conditions, become gravitationally unstable.

We shall focus on collisions between subhalos where both members of the pair are above the atomic-cooling threshold at infall, i.e. have a mass corresponding
to a virial temperature of $T_{\rm vir}>8000$ K,
\begin{equation}
m_{\rm peak}> 1.6\times 10^8\,\msun\, [\mu(1+z)/5]^{-3/2}\,{\cal B}^{-1/2}.
\label{eq:Mc}
\end{equation}
Here, $\mu$ is the mean molecular weight per particle ($\mu=1.23$ for neutral primordial gas), ${\cal B}\equiv \Delta_c/(\Omega_m^z\,18\pi^2)$,
$\Delta_c=18\pi^2+82y-39y^2$ is the redshift-dependent density contrast at virialization \citep{bryan98}, $y\equiv \Omega_m^z-1$,
$\Omega_m^z=\Omega_m(1+z)^3/[\Omega_m(1+z)^3+\Omega_\Lambda]$, and $z$ is the infall redshift. In evaluating these expressions, we have
assumed a \citet{planck18} flat cosmology with parameters $(\Omega_m, \Omega_\Lambda, h)= (0.315, 0.685, 0.674)$, and ${\cal B}=1$. 
Atomic cooling subhalos are likely to be polluted by metals through inefficient star formation, which is key   
since GCs will inherit the gas-phase metallicity of the colliding subhalo population.  
According to cosmological radiation hydrodynamics simulations by \citet{wise14} that follow the buildup of dwarf galaxies from their early
Population III progenitors (see also \citealt{ricotti16}), most atomic-cooling halos host metal-enriched stars by redshift 6, when infalling 
substructures begin colliding at high speed. The gas shock heated in a close encounter will then be pre-enriched by previous generation of massive star formation.

Gas distributed isothermally in atomic-cooling halos develops warm central cores of constant baryon density 
$n_c\approx 0.6\,[(1+z)/5]^3\,\cc\equiv n_1\,\cc$ 
within radii $r_c\sim 0.1\,r_{\rm vir}$ (roughly independent of halo mass) \citep{ricotti09,prieto13,visbal14}, 
and follows an $r^{-2}$ profile in the outer regions. Here,
\begin{equation}
r_{\rm vir}\simeq 4.4\,{\rm kpc}\,\left({m_{\rm peak}\over 10^{8.5}\,\msun}\right)^{1/3}\,\left({5\over 1+z}\right)
\end{equation}
is the virial radius of the system. 
%
%This is comparable to the median impact parameter of our close collision sample (see Fig. \ref{fig3}).
%
Prior to infall, the susceptibility of such systems to internal and external feedback mechanims
may decrease their gas content to 5-10\% of their virial mass \citep{wise14}. After infall, their centrally concentrated gaseous cores
are expected to survive disruption by ram pressure before the first pericenter passage \citep{mayer06,visbal14b}.

\subsection{Shock Heating, Cooling, and Gravitational Instability}
\label{sec:instability}

\begin{figure}
\plotone{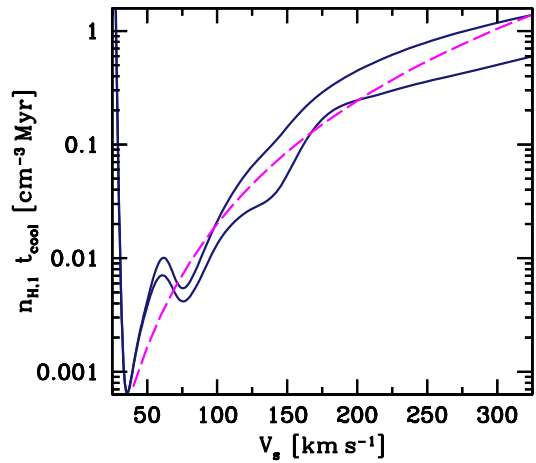}
\caption{Radiative cooling timescale $t_{\rm cool}$ times the pre-shock hydrogen density $n_{\rm H,1}$, as a function of shock speed for atomic gas with
metallicity $Z=0.01\,Z_\odot$ (top curve) and $Z=0.1\,Z_\odot$ (bottom curve). The cooling function and electron 
fraction in collisional ionization equilibrium have been taken
from \citet{wang14}. Strong shock jump conditions have been assumed. The power-law approximation used in Equation (\ref{eq:tcool})
for low metallicity gas, is plotted with the dashed line in the range $40<V_s<300\,\kms$.
\label{figcool}}
\vspace{+0.5cm}
\end{figure}

To highlight the dominant physical processes at work during a high-speed close interaction,
let us consider the idealized situation of a head-on collision between two identical gaseous cores of characteristic hydrogen number density $n_{\rm H,1}=0.75n_c$, mass
density $\rho_1=n_cm_p$, temperature $T_1=8000$ K, and radius $r_c=0.1\,r_{\rm vir}$. These are the density and thermal pressure conditions of the warm 
diffuse component of the interstellar medium. In the absence of molecular cooling, 
even when gas is heated by shocks and compression 
above the $8000\,$K limit, atomic radiative losses together with photoelectric heating from dust grains, cosmic ray heating, and soft X-ray heating will keep the gas
temperature near this value \citep{wolfire95}. The total gas mass involved in the impact,
\begin{equation}
\begin{split}
2m_c= & {8\pi\over 3} \rho_1 r_c^3\simeq 10^{7}\msun ~n_1\left({r_c\over 0.1\,r_{\rm vir}}\right)^3\times \\
& m_{8.5}\left({5\over 1+z}\right)^3,
\label{eq:mcore}
\end{split}
\end{equation}
where $m_{8.5}\equiv m_{\rm peak}/10^{8.5}\,\msun$, is about 10 per cent of the mass corresponding to the universal baryon fraction. 
Following a highly supersonic interpenetrating impact, most of the gas decouples from the collisionless component. Two shock waves arise 
and propagate from the contact discontinuity with velocity (in the adiabatic phase) $V_s\simeq 2v_{\rm rel}/3$ relative to the unpertubed medium, heating the gas to 
\begin{equation}
T_2={3\mu m_p\over 16k_B}V_s^2 \simeq 1.3\times 10^5\,{\rm K}~ V^2_{100},
\label{eq:Tshock}
\end{equation}
where $V_{100}\equiv V_s/100\,\kms$ and we have set the molecular weight to $\mu=0.59$. The density is enhanced by a factor 4, the post-shock 
pressure is $p_2=(3/4)\rho_1V_s^2$, and the gas cools via hydrogen, helium, and metal line emission on a (isobaric) timescale
\begin{equation}
t_{\rm cool}={3k_BT_2n_{\rm tot}\over 2n_{\rm H,2}^2\Lambda} \simeq 0.03\,{\rm Myr}~ \left({V_{100}^{3.6}\over n_1}\right).
\label{eq:tcool}
\end{equation}
Here, $n_{\rm tot}$ is the total particle number density ($n_{\rm tot}/n_{\rm H,2}=9/4$ for completely ionized gas of primordial composition),
$n_{\rm H,2}=4\,n_{\rm H,1}$ is the post-shock mean hydrogen density, $\Lambda$ is the cooling function \citep[e.g.,][]{wang14}, 
and we have assumed collisional ionization equilibrium and ignored gas clumping (see Fig. \ref{figcool}).
% and $C=\langle n^2_{\rm H,2}\rangle/n^2_{\rm H,2}$ is a clumping factor that accounts for gas inhomogeneities on small scales. 
Since $t_{\rm cool}$ is typically much shorter than the crossing time,
\begin{equation}
t_{\rm cross}\equiv {2r_c\over V_s}\simeq 7.8\,{\rm Myr}~ (V_{100}^{-1}\,r_{0.4}),
\label{eq:tcross}
\end{equation}
where $r_{0.4}\equiv r_c/0.4\,{\rm kpc}$,
the internal energy increase caused by the shock is radiated away and the impact is effectively isothermal ($T_2=T_1=T$), characterized by a 
sound speed $c_s=\sqrt{k_BT/\mu m_p}$ and a Mach number
\begin{equation}
{\cal M}\equiv {V_s\over c_s} \simeq 13.6~ (T_{3.9}^{-1/2}\,V_{100}),
\label{eq:Mach}
\end{equation}
where $T_{3.9}\equiv T/8000\,{\rm K}$ and we have assumed $\mu=1.23$. The post-shock hydrogen density is 
\begin{equation}
n_{\rm H,2}={\cal M}^2\,n_{\rm H,1}\simeq 140\,{\rm cm^{-3}} ~(n_1\,T_{3.9}^{-1}\,V^2_{100}),
%{\cal M}_{13.6}^2, 
\label{eq:psdensity}
\end{equation}
and the shock velocity relative to the unpertubed medium is 
now $V_s\simeq v_{\rm rel}/2$. The collision is also approximately one-dimensional, since the lateral rarefaction timescale is 
$\sim {\cal M}\,t_{\rm cross}$. For a strictly isothermal collision, when the compression phase is completed after a time $t_{\rm cross}$, a rarefaction wave 
will propagate through the shock-compressed 
gas slab at the local sound speed, and the medium will rapidly expand into the surroundings on a timescale $\sim t_{\rm cross}/{\cal M}$.
If the collision is to result in gravitational instability and collapse, there must be density perturbations that can grow on a timescale $\lta t_{\rm cross}$. 
We shall see below, however, that gas further downstream will actually cool below its pre-shock temperature via fine-structure metal lines, i.e.  
{the collision is ``more dissipative than isothermal"} \citep[e.g.][]{whitworth97}.  

\Lya\ cooling becomes inefficient below 8000 K, and the fast shock will dissociate most pre-existing molecules. 
In the absence of an ambient UV radiation field, enough molecular hydrogen may reform behind the shock via the H$^-$ channel 
or on the surface of dust grains to rapidly cool the compressed gas to 100 K. It seems likely, however, that photodissociation by
local stellar sources of Lyman-Werner photons -- either within the colliding subhalos or in in the main host -- will act to suppress H$_2$ formation in the post-shock gas, and the 
temperature will then plateau around 8000 K \citep[e.g.,][]{kang90}.  The Bonnor-Ebert mass in such warm, dense, and pressurized medium is
\begin{equation}
\begin{split}
M_{\rm BE} = &  1.182\,{c_s^3\over \sqrt{G^3\,{\cal M}^2\,\rho_1}}\\
\simeq & 7.7\times 10^{5}\,\msun\,(n_1^{-1/2}\,T_{3.9}^{2}\,V_{100}^{-1}).
%M_J = & {c_s^3\over 6}\,\left({\pi^5\over G^3\,{\cal M}^2\,\rho_1}\right)^{1/2}\\
%\simeq & 10^{6.28}\,\msun ~\,n_1^{-1/2}\,T_{3.9}^{3/2}\,{\cal M}_{13.6}^{-1}, 
\label{eq:Jeans}
\end{split}
\end{equation}
%
%where ${\cal M}_{13.6}\equiv {\cal M}/13.6$. 
a factor of $1/{\cal M}$ smaller than its value in the pre-shock gas (where $M_{\rm BE}>m_c$, i.e. the pre-collision clumps are gravitionally stable), 
and comparable to the observed mass of GCs.

GCs are generally metal poor. While the lowest metallicity globular listed in the 2010 edition of the \citet{harris96} catalog has [Fe/H]=$-2.4$ dex, 
the median of the Galactic GC population lies at [Fe/H]=$-1.3$ dex. 
%lowest and upper quartiles are -1.7 and -0.75
Even at the low metallicities of blue globulars, however, the post-shock dense medium will 
eventually cool below 8000 K via the collisional excitation of the fine-structure lines of \CII\ (158 $\mu$m) and \OI\ (63 $\mu$m). The rate coefficient
for the collisional excitation of \OI\ by H atoms is $k_{\rm OI}\simeq 1.3\times 10^{-9}\,{\rm cm^3\,s^{-1}}\,T_{3.9}^{0.41}e^{-T_{\rm OI}/T}$
where $T_{\rm OI}=228$ K is the energy of the excited \OI\ level over $k_B$ \citep[][]{draine11}. At densities far below the critical density of the line, this process
removes energy at the rate
\begin{equation}
n_{\rm H,2}^2\Lambda_{\rm O}=n_{\rm H,2}^2\,{\cal A}_{\rm O}\,k_{\rm OI}(k_BT_{\rm OI}),
\label{eq:LOI}
\end{equation}
where ${\cal A}_{\rm O}$ is the abundance of O relative to hydrogen, ${\cal A}_{\rm O}=4.3\times 10^{-5}\,Z_{-1.3}$ \citep{anders89}, 
$Z_{-1.3}$ is the gas metallicity in units of $10^{-1.3}$ solar, and all oxygen is assumed to be neutral and in the gas phase. 
The isobaric cooling timescale,
\begin{equation}
t_{\rm cool}^{\rm O}={5k_BT n_{\rm tot}\over 2n_{\rm H,2}^2\Lambda_{\rm O}}\simeq 0.4\,{\rm Myr} ~\left({T_{3.9}^{1.59}\over V_{100}^2 n_1 Z_{-1.3}}\right),
%t_{\rm cool}^{\rm O}={5k_BT n_{\rm tot}\over 2n_{\rm H,2}^2\Lambda_{\rm O}}\simeq 0.4\,{\rm Myr} ~{T_{3.9}^{0.59}\over {\cal M}_{13.6}^2 n_1 Z_{-1.3}},
\label{eq:tcoolO}
\end{equation}
is still shorter than the crossing time, but it is considerably longer than $t_{\rm cool}$ in Equation (\ref{eq:tcool})
as a result of the mismatch in timescales between fine-structure metal-line cooling below 8000K and atomic cooling at higher temperatures.  After a time $\sim t_{\rm cool}^{\rm O}$, 
the flow will cool below its pre-shock temperature and contract further, generating a growing, cold and very dense layer at the center of the shock-bounded slab.

\begin{figure}
\plotone{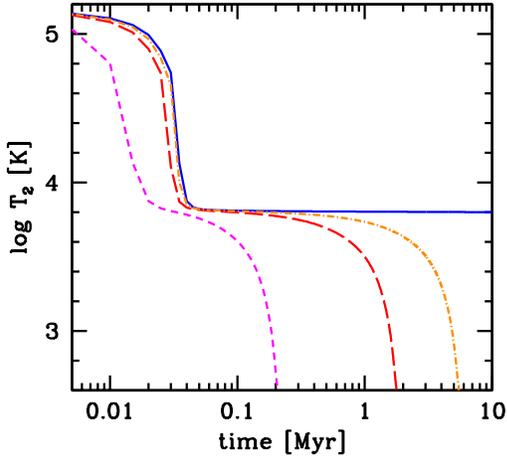}
\caption{Post-shock temperature vs. time (in Myr, measured since the fluid element was shocked) for a case with $V_s=100\,\kms$ and $n_{\rm H,1}=0.75\,\cc$.
Strong shock jump conditions and isobaric cooling have been assumed, and the non-equilibrium chemistry network has been  
taken from the {\sc krome} package \citep{grassi14}. The gas is also assumed to be purely atomic, dust-free, and optically thin.
The thermal model includes atomic cooling, metal cooling following \citet{bovino16}, and $pdV$ work.
Four curves are shown for different metallicities: $Z=0$ (solid line), $Z=10^{-2.5}\,Z_\odot$ (dot-dashed line), $Z=10^{-2}\,Z_\odot$ (long-dashed line),
and $Z=10^{-1}\,Z_\odot$ (short-dashed line).
%The dotted vertical line shows the crossing time $t_{\rm cross}=2r_c/V_s$ for $V_s=100\,\kms$ and $r_c=0.4\,$kpc.
\label{figKrome}}
\end{figure}

The arguments above are based on simple timescale estimates, and to check the robustness of our results we have used the non-equilibrium 
chemistry package {\sc krome} \citep{grassi14} to construct more sophisticated models of the cooling and thermal properties of the post-shock gas. 
The thermal model includes non-equilibrium cooling for H and He at all temperatures and for metals below $10^4$ K \citet{bovino16}, and compressional heating.
Strong shock jump conditions and isobaric cooling were adopted for gas at different metallicities (ranging from 0 to 0.1 solar), and the cooling function 
was computed assuming dust-free and optically thin conditions, and neglecting molecular-phase processes. Figure \ref{figKrome} shows 
the time evolution of the post-shock temperature for a case with $V_s=100\,\kms$ and $n_{\rm H,1}=0.75\,\cc$.
The figure is consistent with the qualitative estimates given above: low-density gas, shock-heated to high temperatures, 
cools fast via atomic hydrogen and helium line emission to about 6500 K, and then levels off for a timescale that is longer at lower metallicities.
Eventually, metal cooling via fine-structure lines of C and O takes over. The end result is a runaway cooling phenomenon that drives the shocked 
medium to temperatures $T_3<T$ and hydrogen densities $n_{\rm H,3}=n_{\rm H,2}T/T_3$ (assuming isobaric cooling at constant mean molecular weight) that 
are, respectively, well below and above those of the colliding gas clouds. Note that these calculations do not explicitly include internal (within the colliding substructures) 
and external (within the host galaxy) sources of UV and X-ray photoheating. 

The one-dimensional compression will create a cold layer of thickness-to-diameter ratio $\sim 2n_{\rm H,1}/n_{\rm H,3}<2/{\cal M}^2\sim 0.01~(T_{3.9}\,V_{100}^{-2})$ 
that depends on the entity of radiative losses. In this case, gravitational accelerations are most important in the dynamics of flows transverse to the collision 
axis, and it is more appropriate 
to discuss the criterion for the gravitational instability of a thin, ram pressure bounded slab of shocked gas \citep[e.g.,][]{stone70,elmegreen78,vishniac83,gilden84,larson85}. 
For an infinitely thin sheet of constant surface density $\Sigma$ and {\it isothermal} sound speed $c_s$, the dispersion relation for modes transverse
to the collision axis gives the wavelength and timescale of the fastest growing mode as
\begin{equation}
\lambda_G={2c_s^2\over G\Sigma},
\end{equation}
and
\begin{equation}
t_G={c_s\over \pi G\Sigma}.
\label{eq:tG}
\end{equation}
Perturbations smaller than $\lambda_G$ are gravitationally stable, while those with longer wavelengths are unstable but grow more slowly.
The surface density of the cold layer between the two shocks increases with time as $\Sigma(t)=2\rho_1V_st$,
%along the line between the cloud centers, 
where $t$ is measured from the onset of metal cooling below 8000 K. The shocked gas is first liable to gravitational fragmentation at 
time $t=t_G$, when $\Sigma_G\equiv \Sigma(t_G)=2\rho_1V_st_G$ \citep{kang90}. Using Equation (\ref{eq:tG}) one derives then
\begin{equation}
t_G=\left({c_s\over 2\pi G\rho_1 V_s}\right)^{1/2}\simeq 3.8\,{\rm Myr}~\left({{\cal C}\over n_1V_{100}}\right)^{1/2},
\end{equation}
and
\begin{equation}
\Sigma_G=\left({2\rho_1 V_s c_s\over \pi G}\right)^{1/2}\simeq 19\msun {\rm pc^{-2}}~(n_1V_{100}{\cal C})^{1/2},
\end{equation}
where ${\cal C}\equiv c_s/(1\,\kms)$. Non-linear fragmentation occurs before the start of the rarefaction era 
when $t_G\le t_{\rm cross}-t_{\rm cool}^{\rm O}\simeq t_{\rm cross}$. This inequality is fulfilled when 
\begin{equation}
c_s\le {8\pi\,G\rho_1 r_c^2\over V_s}\simeq 4.3\,\kms ~(V_{100}^{-1}\,n_1\,r_{0.4}^2) 
\label{eq:c0}
\end{equation}
(corresponding with the choosen scalings to $T_3\le 2750\,$K). In the case of very fast collisions, the gas must cool to increasingly lower temperatures 
for gravitational instability to act on short enough timescales.  As soon as condition (\ref{eq:c0}) is satisfied, the unstable slab will 
break up into circular fragments of preferred radii 
\begin{equation}
{\lambda_G\over 2}={\pi c_s t_G}\simeq 107\,{\rm pc}~\left({r_{0.4}^3 n_1\over V_{100}^2}\right) 
\label{eq:lG}
\end{equation}
and masses
\begin{equation}
{M_G}= {\pi\Sigma_G}\,\left({\lambda_G\over 2}\right)^2 \simeq 10^{6.15}\,\msun ~\left({r_{0.4}^7 n_1^3\over V_{100}^4}\right).
\label{eq:MG}
\end{equation}
While the latter is again comparable to the observed characteristic mass of GCs after accounting for mass loss by stellar evolution and tidal disruption after
birth,\footnote{Note that some scenarios for the formation of multiple stellar populations within GCs require globulars to have been initially much more 
massive than they are today, \citep[e.g.][]{dercole08}.}\, it is also somewhat ill-defined -- scaling strongly with cloud size, gas density, and shock velocity. 
Neverthless, it is encouraging that collisions between atomic cooling, metal-poor subhalos may offer a mechanism for imprinting the 
signature of the GC mass scale on the collapsing shell. 

It should be noted that the conditions for gravitational instability given in Equations (\ref{eq:c0}-\ref{eq:MG}) can only be met 
as long as the gas metallicity $Z$ is not much below $10^{-2}\,Z_\odot$ (see Fig. \ref{figKrome}). This may provide a plausible 
explanation for the threshold metallicity, [Fe/H]$=-2.4$ dex, below which GCs are not observed. 
It is also significant that, at the time when gravitational instability sets in, the gas will be crushed into 
a slab of thickness $d$  along the line connecting the cloud centers, 
\begin{equation}
d={4r_cT_3\over {\cal M}^2 T} \simeq 3\,{\rm pc}~(r_{0.4}^5\,n_1^2\,V_{100}^{-4}),
\end{equation}
which is comparable to the typical half-light radius of GCs. The free-fall time for a uniform, pressure-free sphere of such gas, 
\begin{equation}
\begin{split}
t_{\rm ff} = & \sqrt{{3\pi\over 32 G\rho_3}}={1\over {\cal M}}\,\sqrt{{3\pi T_3\over 32 G \rho_1 T}}\\
\simeq & 2.2\, {\rm Myr}\,(V_{100}^{-2}n_1^{1/2}\,r_{0.4}^2),
\end{split}
\end{equation}
is shorter than the several-Myr evolutionary timescale for massive stars.

The above analysis is highly idealized, and is only meant to provide an idea of the general conditions under which the splash remnant may become Jeans 
unstable. It is clear from the previous discussion that one expects significant spatial structures in the post-shock region both in temperature and density as a function
of distance from the shock fronts \citep[e.g.,][]{smith80,kang90,kumai93}, and the thin shell approximation may be inadequate in the
case of very inhomogeneous flows \citep{yamada98}. A shocked slab is known to be susceptible to a number of hydrodynamical instabilities like the non linear thin shell 
instability \citep{vishniac94}, which may compete with the gravitational instability and produce substructures on the scale of the slab thickness.
The thermodynamic treatment of the problem has been simplified by neglecting molecular cooling, heating from external X-ray radiation and cosmic rays, dust processes, and
gas clumping.
%as well as the possible feedback effects from the massive stars born in the shock-compressed region. 
The simple model of head-on collisions between identical, uniform-density clumps is obviously unrealistic, and should be extended to off-center impacts 
between subhalos with internal substructure. 

Nevertheless, our calculations may elucidate the conditions for a special, dynamical mode of star formation following substructure collisions, a mode that 
is intimately tied to $\Lambda$CDM hierarchical galaxy assembly. 
It seems plausible that GC formation by impact requires the relative velocities of the colliding subhalos to be in a specific 
range. If the collision velocity is too low, shocks may not be able to produce the extremely high pressure environments that are a prerequisite to the formation of dense and 
tightly bound clusters. Conversely, if the velocity is too high and the shock too violent, the interacting clouds will expand and disperse before significant 
radiative cooling can occur. Equations (\ref{eq:tcool}) and (\ref{eq:tcross}) give
\begin{equation}
{t_{\rm cool}\over t_{\rm cross}}\simeq 3.5\times 10^{-3}\,\left({V_{100}^{4.6}\over n_1\,r_{0.4}}\right),
\end{equation}
and $t_{\rm cool}/t_{\rm cross}<1$ for $V_s<343\,\kms ~(n_1\,r_{0.4})^{0.217}$.
This upper velocity limit is only weakly dependent on the properties (gas density and size) of the original colliding clumps.

Simulations of interstellar cloud-cloud collisions may also provide some insight on the fate of the shock-compressed layer. 
According to \citet{balfour15}, when cold, uniform-density clouds collide head-on at moderately supersonic speeds, 
star formation operates in a global hub-and-spoke mode that produces a central monolithic stellar cluster.
At higher collision velocities, a spider's-web mode operates and delivers a loose distribution of independent, 
small sub-clusters instead. When the clouds have pre-collision substructure, however, the collision
velocity becomes less critical \citep{balfour17}. In these numerical experiments the gas is evolved with a barotropic equation of state. 
Ultimately, the detailed fate of subhalo-subhalo high-speed collisions should be addressed with the help of cosmological 
hydrodynamic simulations that include all the relevant heating and cooling processes.

\section{N-body Simulation} \label{S:VLI}

High-speed close interactions between satellites orbiting within a parent halo have been advocated as a major mechanism for the morphological 
evolution of galaxies in clusters \citep{moore96h,baushev18}, and collisions between protogalaxies have been proposed as a new pathway to form 
supermassive black holes at very high redshifts \citep{inayoshi05}.
Cosmological $N$-body simulations by \citet{tormen98} showed that fast satellite-satellite encounters with impact parameter $b<(r_{\rm vir,1}+r_{\rm vir,2})$
are fairly common and can lead to significant mass loss and disruption. In this section we make use of the {\it Via Lactea I} (VLI) $N$-body simulation to 
argue that rarer, nearly-central collisions between atomic-cooling subhalos still occur frequently enough at high redshifts to 
represent a plausible pathway to the formation of GCs.

%and the Aquarius suite of MW-sized halo simulations \citep{springel08}, %and 3.1, respectively. 
%To correct for this shortfall, we have artificially increased the
%
\begin{figure*}
\epsscale{0.94}
\plottwo{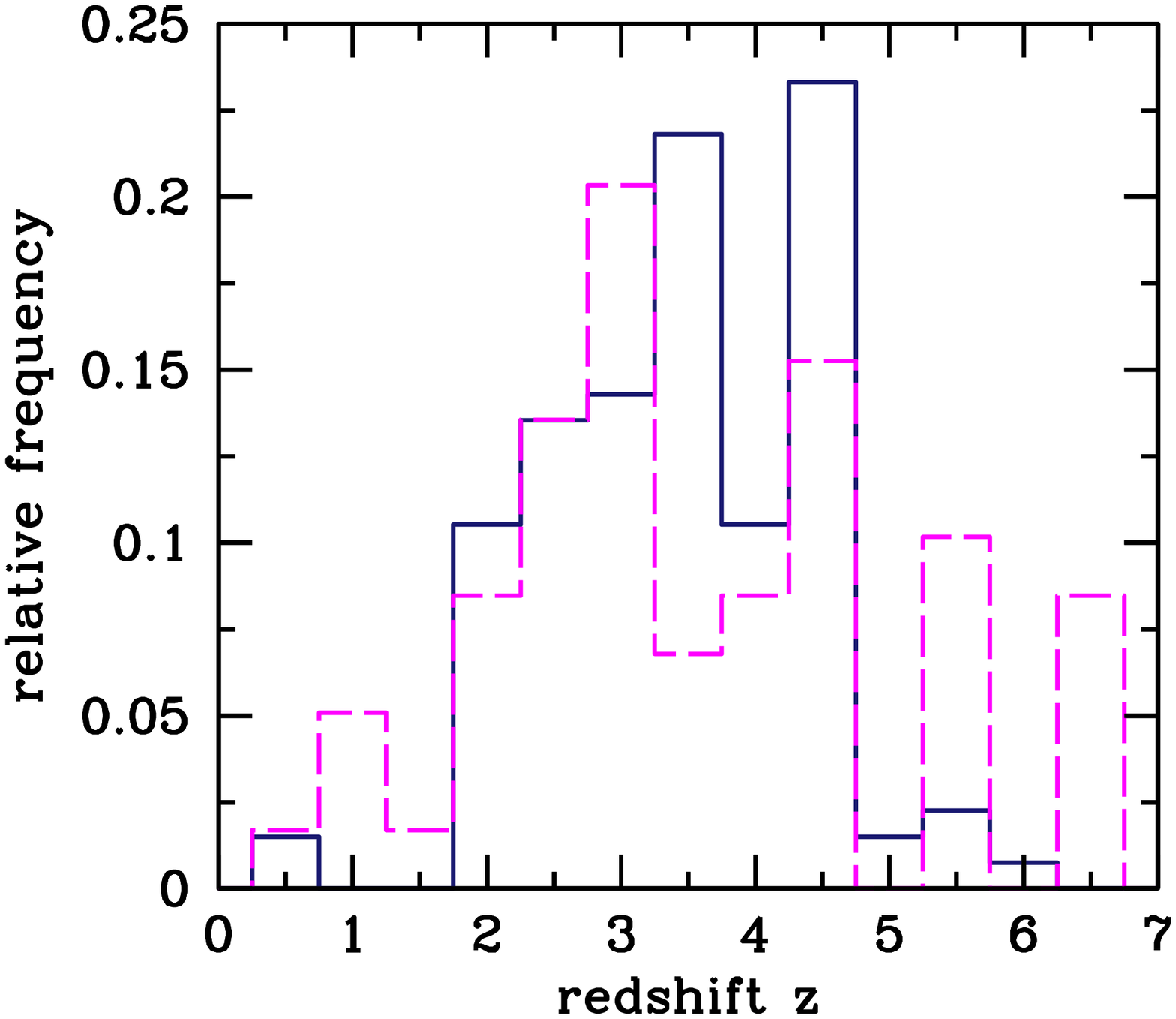}{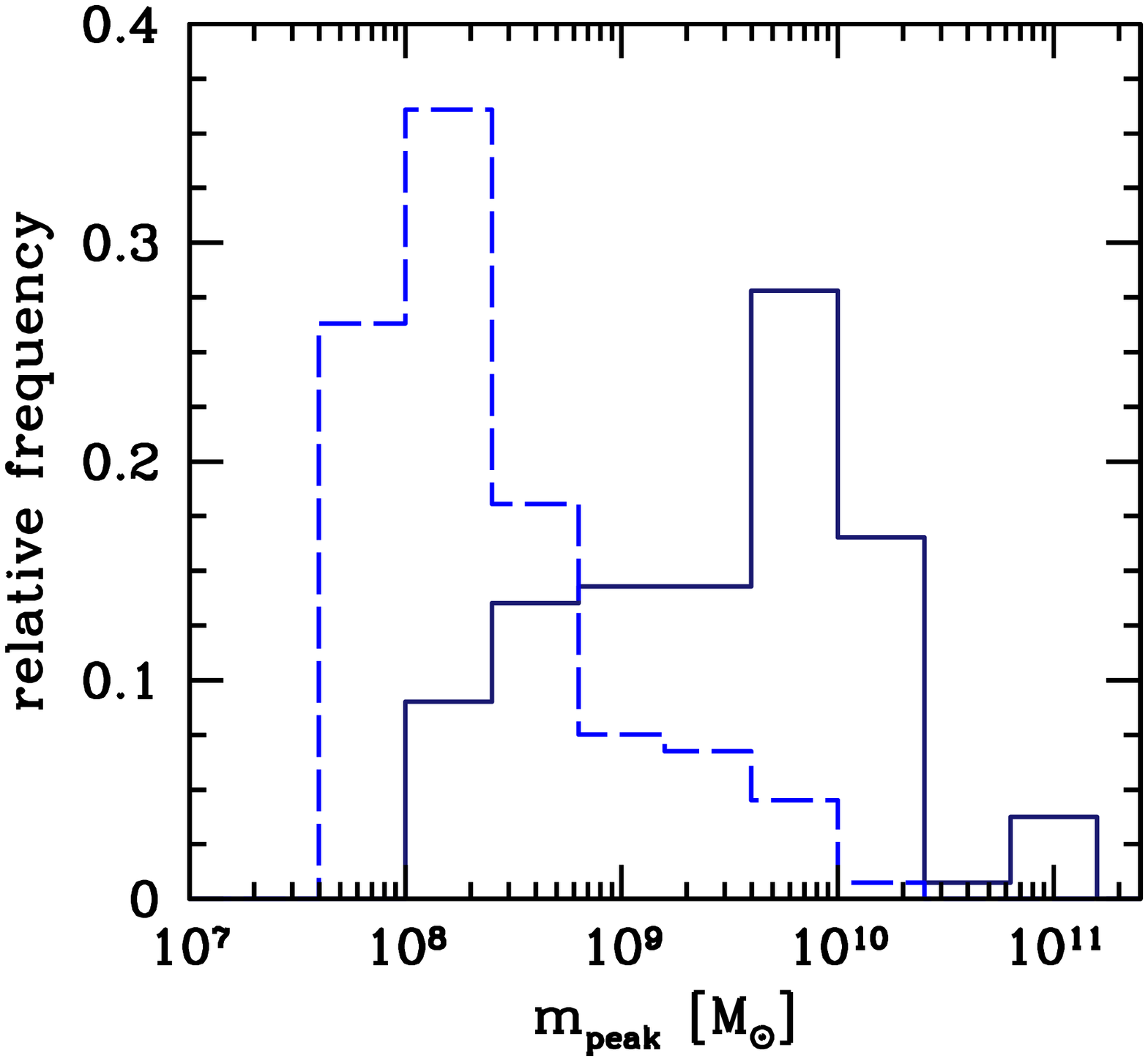}
\plottwo{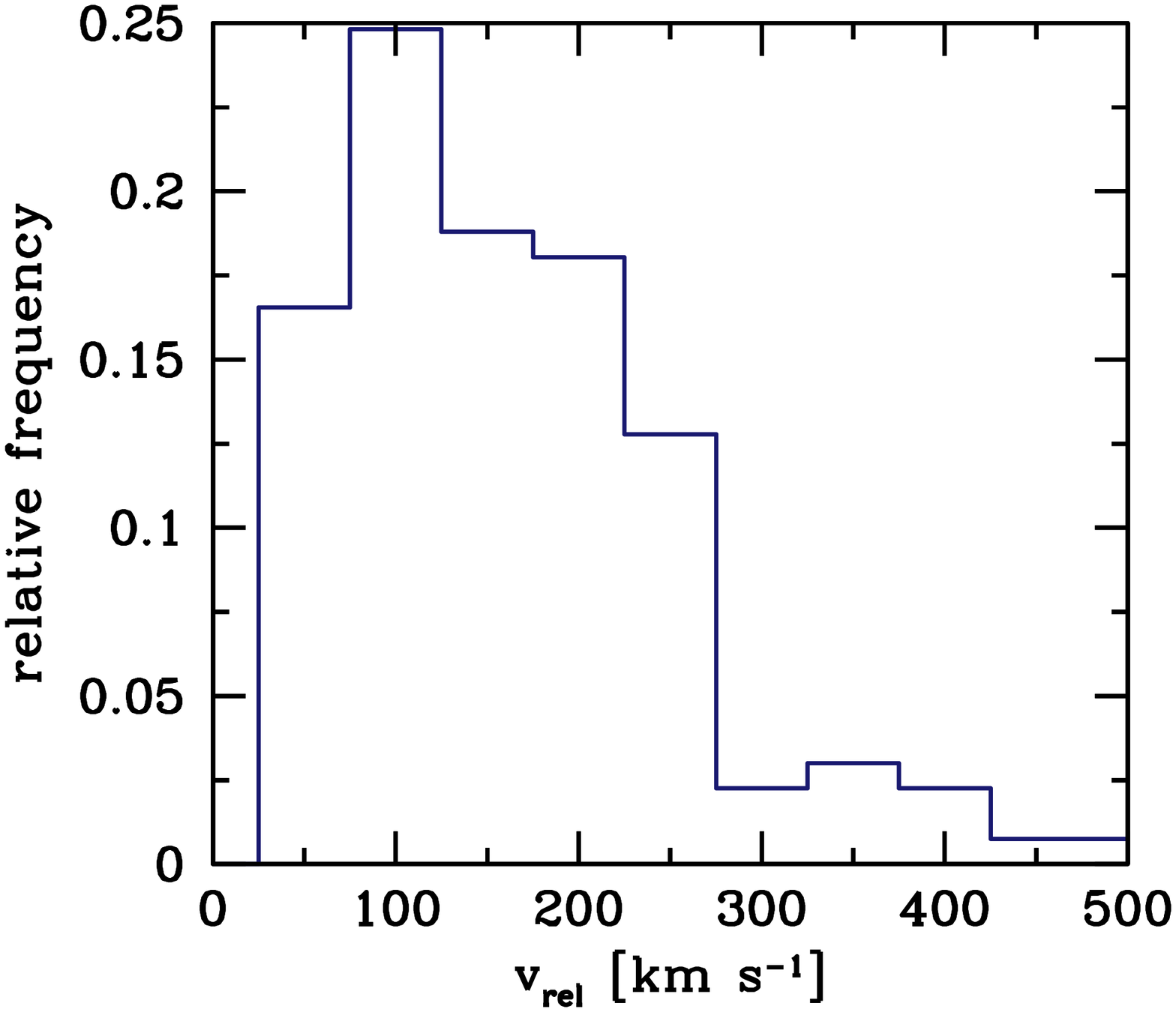}{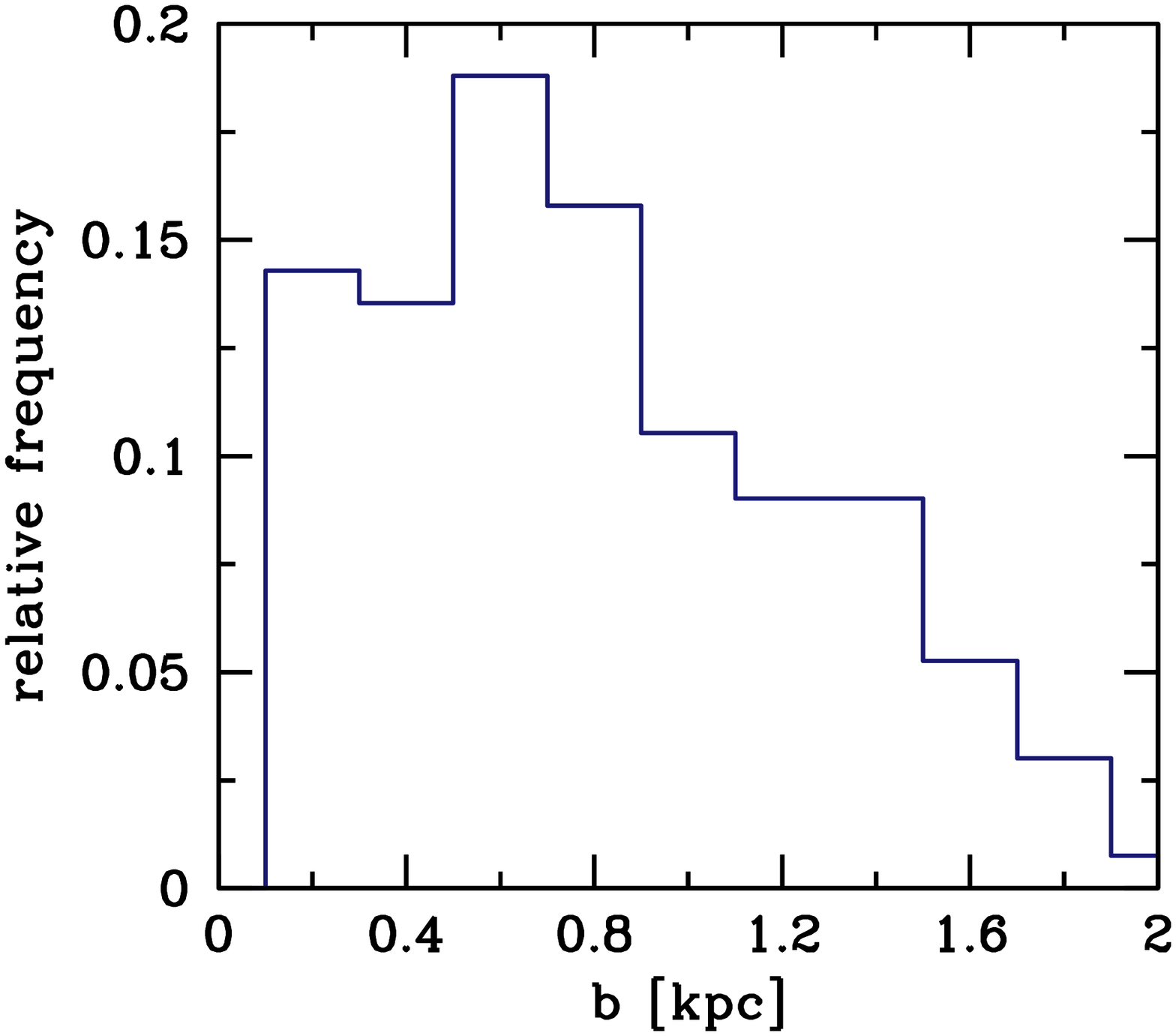}
\plottwo{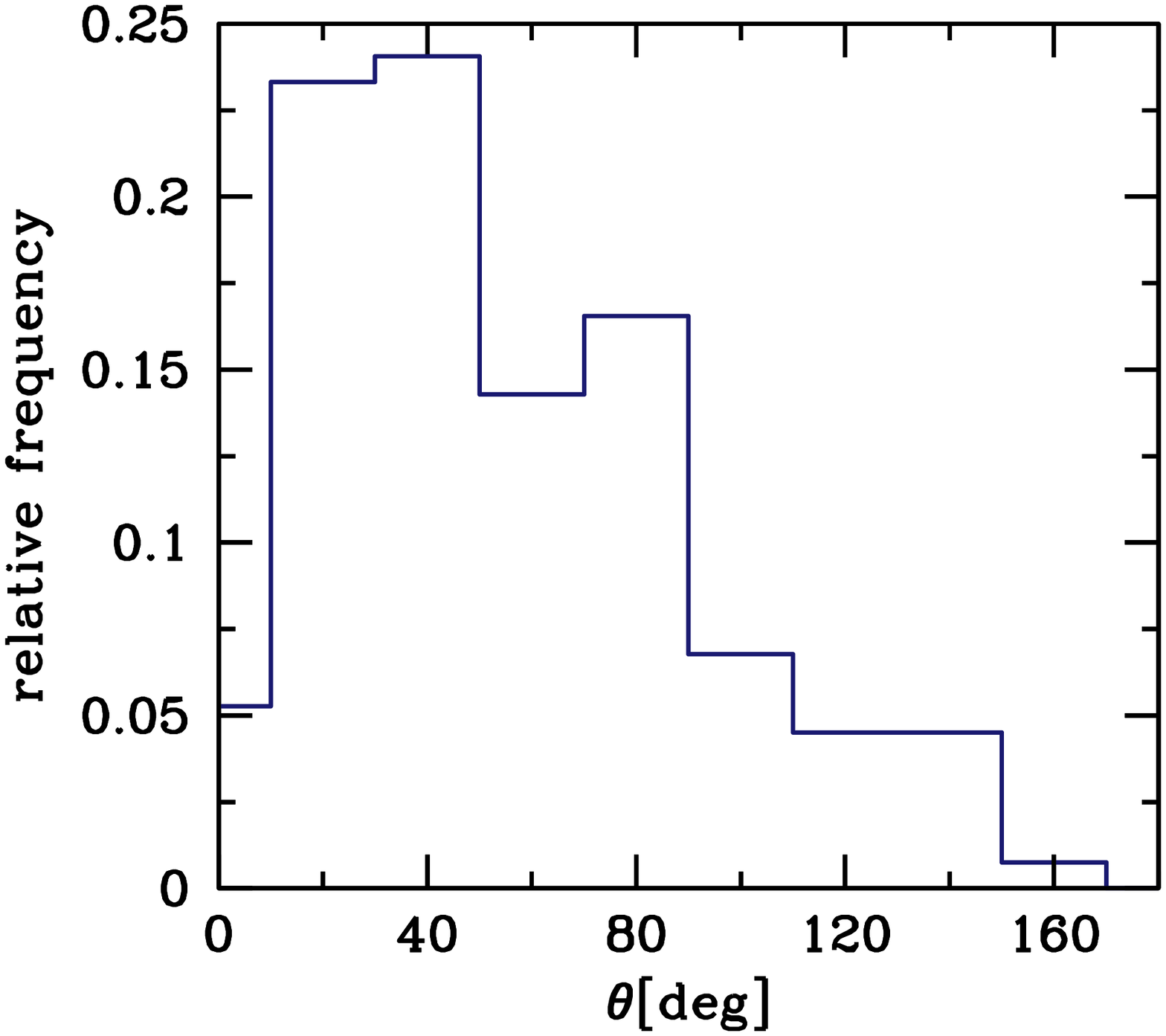}{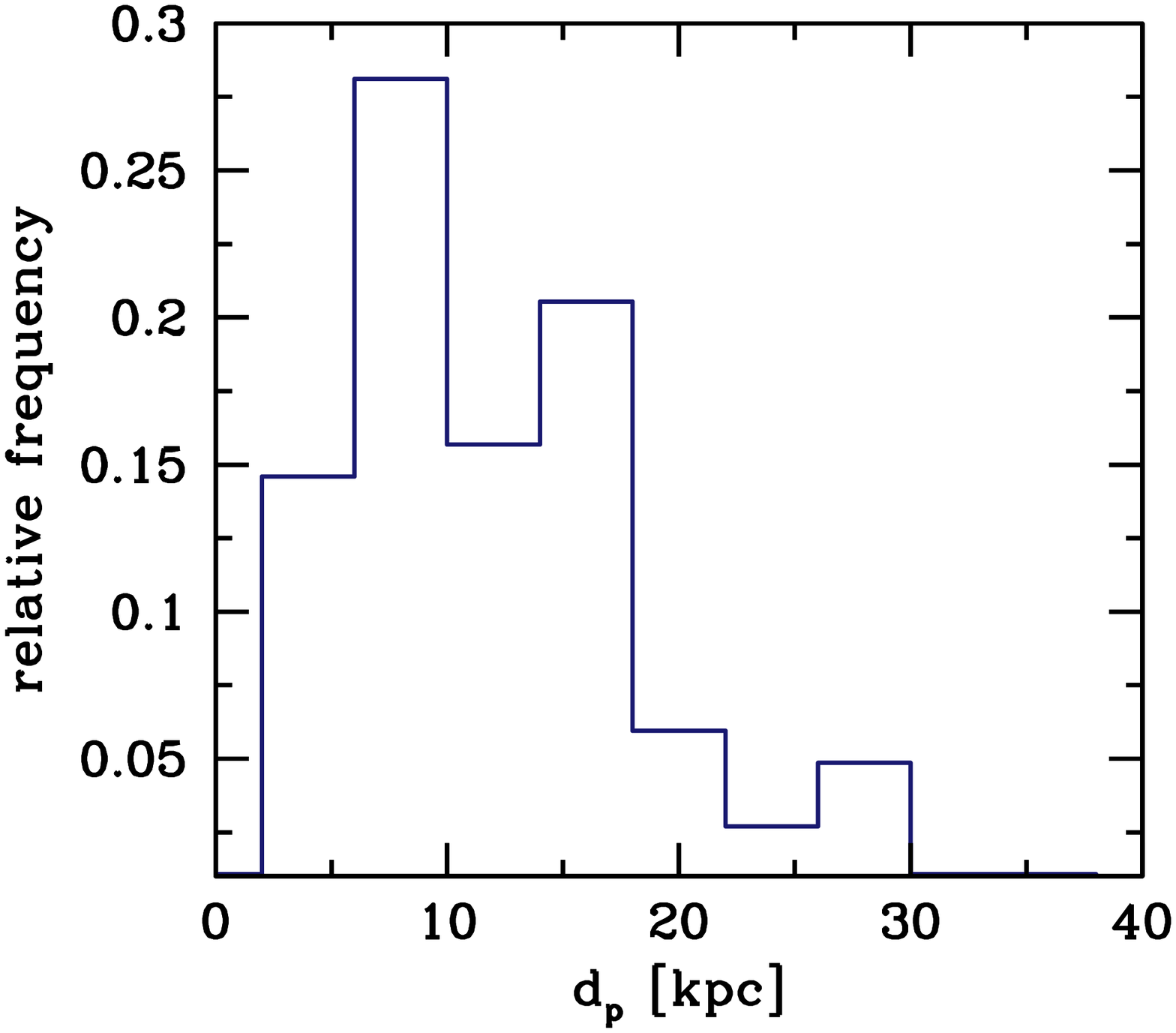}
\caption{Frequency distribution of redshifts (top left panel), peak masses (top right panel),
relative velocities (middle left panel), impact parameters (middle right panel), 
impact angles (bottom left panel), and first pericenter distances (bottom right panel)
for all unbound interpenetrating collisions between atomic cooling subhalos in the {\it Via Lactea} simulation (see text for details). 
The top right panel shows the peak mass frequency histograms for both the lighter (dashed line) and heavier (solid line) members of all colliding pairs, 
while the bottom right panel depics the first pericenter distance for all colliding subhalos that reach pericenter before disruption.
The dashed histogram in the left top panel delineates the redshifts of formation for a sample of Milky Way GCs. The age determinations 
by \citet{VdB13}, \citet{sarajedini07}, \citet{dotter10}, \citet{valcheva15}, and \citet{weisz16} have been converted
to redshift using a \citet{planck18} cosmology. The distribution is poorly known at $z\gta 4$ as the typical
uncertainty on the absolute age of GCs exceeds 1--1.5 Gyr. 
\label{figVLI}}
\end{figure*}

VLI is a dark matter-only simulation that follows the formation of a MW-sized halo %of mass $M_{\rm vir}=1.8\times 10^{12}\,\msun$
in a  $\Lambda$CDM cosmology. The high-resolution region was sampled with 234 million particles of mass $2.1\times 10^4\,\msun$ and
evolved with a force resolution of 90 pc starting from redshift 50 \citep{diemand+07,diemand07}. We stored and analyzed 200 outputs
from redshift 16 to $z=0$. The simulation has sufficient mass resolution to follow metal and atomic-cooling subhalos through many
orbits and severe mass stripping, and sufficient output time resolution ($\Delta t=68.5\,$Myr) to measure the orbital parameters
of subhalos with good accuracy. Such output spacing provides many timesteps from infall to the first pericenter passage per subhalo.
Substructure catalogs were constructed using the phase-space group-finder 6DFOF \citep{diemand06}, subhalos were linked across snapshots
to follow their histories and trajectories backward and forward in time, and all substructure-substructure close encounters were recorded.
When tracing halos backward in time, a subhalo was linked to its main progenitor only if the core of the latter contained at least 50\% of the
particles of the core of the former, and viceversa.\footnote{The phase-space group-finder 6DFOF finds peaks in phase-space density and links the
particles within those peaks into groups. These groups correspond to the cores of halos and subhalos, whose total extent stretches
well beyond these cores and is found by 6DFOF in a second step \citep{diemand06}.}\,

VLI initial conditions were generated with the original version of the GRAFIC2 package \citep{bertschinger01}, which incorrectly used
the baryonic instead of the dark matter power spectrum for the refinement levels, leading to reduced small-scale power. Compared to {\it
Via Lactea II} \citep{diemand08}, the abundance of subhalos in VLI is suppressed by a factor of 1.7, and one should therefore view
the derived collision frequency strictly as a lower limit. This is even more so as even state-of-the-art cosmological simulations 
still suffer from significant overmerging, and many subhalos will be artificially disrupted before a collision by numerical 
resolution effects \citep{vdB18}. The distance of closest approach between subhalos was 
found by linearly interpolating distances amid snapshots.  Once a subhalo is accreted by the main host, its diffuse outer layers are rapidly 
stripped off by tidal forces, with tidal mass losses being more significant for more massive subhalos \citep{diemand07}.
%
%Subhalo survived masses $m_{\rm sub}$ are defined as those within a tidal radius $r_t$ where the host local density is half of the satellite density,   
%
The accurate tracking of the accretion history of substructure allows us to define the epoch when the satellite mass reached a maximum value, denoted here 
as $m_{\rm peak}$, before infall.

%=R\sigma_{\rm sub}/\sqrt{2}\sigma$. 
%This radius is the classical Jacobi limit for an isothermal 
%satellite (of one-dimensional velocity dispersion $\sigma_{\rm sub}$) on a circular orbit of radius $R$ within an isothermal host 
%halo of velocity dispersion $\sigma$. 

We have identified all collisions between atomic cooling systems ($T_{\rm vir}>8000\,$K) that, at the instant of closest approach, are unbound and interpenetrating, 
i.e. involve subhalos with relative velocities $v_{\rm rel}>3\,V_{\rm max,1}$ and whose centers are separated by a distance $b<b_{\rm max}=0.1\,(r_{\rm vir,1}+r_{\rm vir,2})$.  
Here $V_{\rm max,1}$ is the maximum circular velocity of the larger (``1") 
of the two subhalos at impact, $3\,V_{\rm max}$ is the escape velocity from the center of a spherically-averaged NFW density profile, and the choosen maximum 
impact parameter $b_{\rm max}$ corresponds to the sum of the gas core radii of the pair.
%We are also interested in collisions between satellites of similar mass -- defined here as having mass ratios at impact that are 1:10 or larger. 
The impact parameter condition ensures that the region of overlap between clumps colliding off-center is extensive 
and so is the resulting gas splash, and excludes events where small, dense clumps may plow through the tenuous halo of larger 
subhalos without much resistance, i.e. without their gas becoming dislodged. In order to minimize the effect of ram-pressure stripping on the 
gaseous content of interacting substructure, the collisions of interest here are those that occur {\it before} the first pericenter passage of each subhalo.
Clumps with correlated infall histories often undergo multiple collisions: for simplicity, we assume here that the smaller member of a colliding pair depletes its gas 
reservoir without replenishment during the first interpenetrating encounter, and do not count any close interactions it may be involved in afterwards. 

Figure \ref{figVLI} shows the frequency distributions of redshifts, masses, relative velocities, impact parameters, angles of impact, and first pericenter 
distances for all encounters that satisfy the above criteria. We tally 133 collision events within today's VLI virial volume, corresponding to 
$R_{\rm vir}=288\,$kpc (note that VLI virial radius drops below 40 kpc at $z\gta 3$). If most of these encounters were to result in the formation of one or more 
massive GCs, the predicted frequency would be comparable to the $\sim$ 150 globulars observed today in the halo of the MW. If, on the other hand, many more low-mass 
GCs formed initially than is currently observed, 
and were selective destroyed by dynamical processes acting over a Hubble time \citep[e.g.,][]{gnedin97,vesperini98}, then our model would require either the formation
of multiple GCs {\it per impact}, or a higher collision frequency, perhaps involving the more numerous population of $T_{\rm vir}<8000$ K subhalos.\footnote{By way 
of illustration, removing the constraints on the virial temperature of interacting subhalos and counting all multiple encounters would result in nearly 
1000 unbound close collisions.}\, Collisions occur at early times in the transition region between the main host and the field: we count only 10 events 
within a Galactocentric distance of 50 kpc. 
For the most part, these collisions involve subhalos that are already dynamically associated before accretion into the main host, i.e. they are either part 
of the same infalling halo or two separate clumps descending along a filament and organized into small groups with correlated trajectories \citep[e.g.,][]{li08,angulo09}.
In this situation, the distinction between {\it ex-situ} and {\it in-situ} globulars becomes blurred.

The median masses at infall of the lighter and heavier member of the colliding pairs are $10^{8.3}\,\msun$ and $10^{9.7}\,\msun$, respectively.
The median relative velocity is $160\,\kms$, and there is a weak negative correlation (with correlation coefficient $r=-0.3$)
between relative velocity and redshift: the few extreme-velocity impacts with $v_{\rm rel}\gta 350\,\kms$ all occur at $z<3.5$, when the depth of the
gravitational potential of the main host is larger.  
%
%Sample size: 133
%Mean x (redshift): 3.4535338345865
%Mean y (rel velocity): 165.51345864662
%Intercept (a): 266.60614123493
%Slope (b): -29.272243281908
%Regression line equation: y=266.60614123493-29.272243281908x
%
Collisions have typical impact parameters in the range $0.3\lta b\lta 1.7\,$ kpc (median value $0.8\,$ kpc), 
and the angles of impact between the initial velocity vectors of the two bodies is less than 50 degrees in about half of all encounters.     
Most interacting subhalos have highly radial orbits and plunge deep into their host halo \citep[see also][]{wetzel11,gonzalez16}. 
In about 40\% of all collisions, one or both of the interacting subhalos are tidally disrupted before falling within 50 kpc of the center of the main host.  

%While a few massive satellites experience multiple collisions, most small subhalos 
%suffer no interpenetrating interactions. 
%In a MW-sized host, where the relative orbital velocities between subhalos substantially exceed their internal velocities, 
%most close encounters are unbound collisions between satellites that are accreted at similar times and locations, and occur in the outer regions of the host halo. 

\section{Summary and Implications for GC Formation Models}

The presence of substructure within dark matter halos is a unique signature of a Universe where systems grow hierarchically through the accretion of 
smaller-mass units. We have investigated a scenario where the formation of GCs is triggered  by high-speed collisions between infalling, atomic-cooling
subhalos during the early assembly of the main galaxy host. This is a special, dynamical mode of star formation that operates at extremely high gas pressures, is 
relatively immune from the feedback processes that regulate star formation in the field, and tracks the total mass of the main host rather than its stellar mass. 
The proposed mechanism would give origin to dark-matter-less globulars, as colliding DM subhalos and their stars will simply pass through one another while the warm 
gas within them shocks and decouples. The well-known Bullet galaxy cluster provides a striking illustration of the process, albeit on 
different scales. In a MW-sized host, where the relative orbital velocities between subhalos substantially exceed their internal velocities, 
most close encounters are unbound collisions between satellites that are accreted at similar early times and locations, and occur in the outer regions of the main 
host. 

Below, we summarize our main results and discuss some implications for GC formation models.

\bigskip {\it (i) Fragmentation of Shocked Gas and the Masses of GCs:}
Shock heating and cooling, the encounter geometry, and the complexities of multiphase gaseous inner halos are all key factors in determining the outcome of a
subhalo-subhalo impact. We have shown that, under idealized conditions, the low metallicity warm gas in the cores of interacting subhalos will be shock-heated 
to characteristic temperatures $\sim 10^5$ K, and will cool rapidly first via atomic line emission and, further downstream, via fine-structure lines of C II and O I. 
Because the collision is more dissipative than isothermal, the resulting shock-compressed slab is liable to gravitational instability.
An idealized analysis in the thin shell approximation yields the conditions under which the imprint of the GC mass scale could be present on the cooling, collapsing shell.
The requirements for gravitational fragmentation can only be satisfied for gas metallicities $Z\gta 10^{-2.5}\,Z_\odot$, thus offering a natural interpretation for 
the observed minimum metallicity of GCs.
We caution, however, that these findings are based on a simplified thermodynamic treatment of the problem that neglects molecular cooling, 
heating from external X-ray radiation and cosmic rays, dust processes, and gas clumping.

\bigskip {\it (ii) The GC system-DM Connection:}
An analysis of the scaling behavior of the encounter frequency within the kinetic theory approximation points to a GC number-halo mass relation that is the result of 
both encounter probability calculations --  the subhalo collision rate per unit volume being the usual density$^2\times$cross section$\times$velocity 
factor -- and of hierarchical assembly -- globulars being brought in by the accretion of smaller satellites. With the addition of {\it ex-situ}  globulars,
the scaling of the total number of GCs with host virial mass is very close to linear, $N_{\rm GC}\propto M_{\rm vir}^{0.98}$, in agreement 
with the trend observed over five orders of magnitude in galaxy mass. This uniform GC production rate per unit host halo mass is predicted
to break down on dwarf galaxy scales, perhaps below a critical mass of $\sim 10^{9.6}\,\msun$.  

%\bigskip {\it (iii) Rate of Cluster Formation:}

\bigskip {\it (iii) The Ages of GCs:}
Our model differs for much previous work as it does not assume an arbitrary value for the redshift when metal-poor GC formation is shut-off. 
The details of the redshift distribution in top left panel of Figure \ref{figVLI} reflect the mass assembly history of the simulated MW-sized host system,
but it is again noteworthy that a scenario in which GCs are the result of colliding substructures would produce a population of
old clusters with typical ages $>10\,$Gyr, a median age of 12 Gyr (corresponding in the adopted cosmology to a median redshift of 3.5), and an age spread that is 
similar to the one observed. In contrast to many pregalactic scenarios \citep[e.g.,][]{katz13,kimm16,boylan17}, in our model GCs have extended formation histories and typically 
form {\it after} the epoch of reionization: only about 38\% of all close encounters occur at redshifts greater than 4. 
Seven collision events in our sample take place at $z<2$. Four of these ``late" impacts have relative velocities in excess of $350\,\kms$, a situation that may not
be conducive to the cooling and fragmention of the splash remnant (see \S~\ref{sec:instability}). The others may give origin to a population of young metal-poor globulars 
like Crater \citep{weisz16}. The dot-dashed histogram on the same panel shows the relative frequency histogram of the redshift of formation for 
55 Milky Way GCs with {\it Hubble Space Telescope} 
photometry \citep{VdB13}, augmented by the age determinations for the young globulars Crater, Pal 1, Terzan 7, and Whiting 1 \citep{sarajedini07,dotter10,valcheva15,weisz16}.
The distribution is poorly known at $z>4$ as the typical uncertainty on the absolute age of GCs exceeds 1--1.5 Gyr.

\begin{figure*}
\plottwo{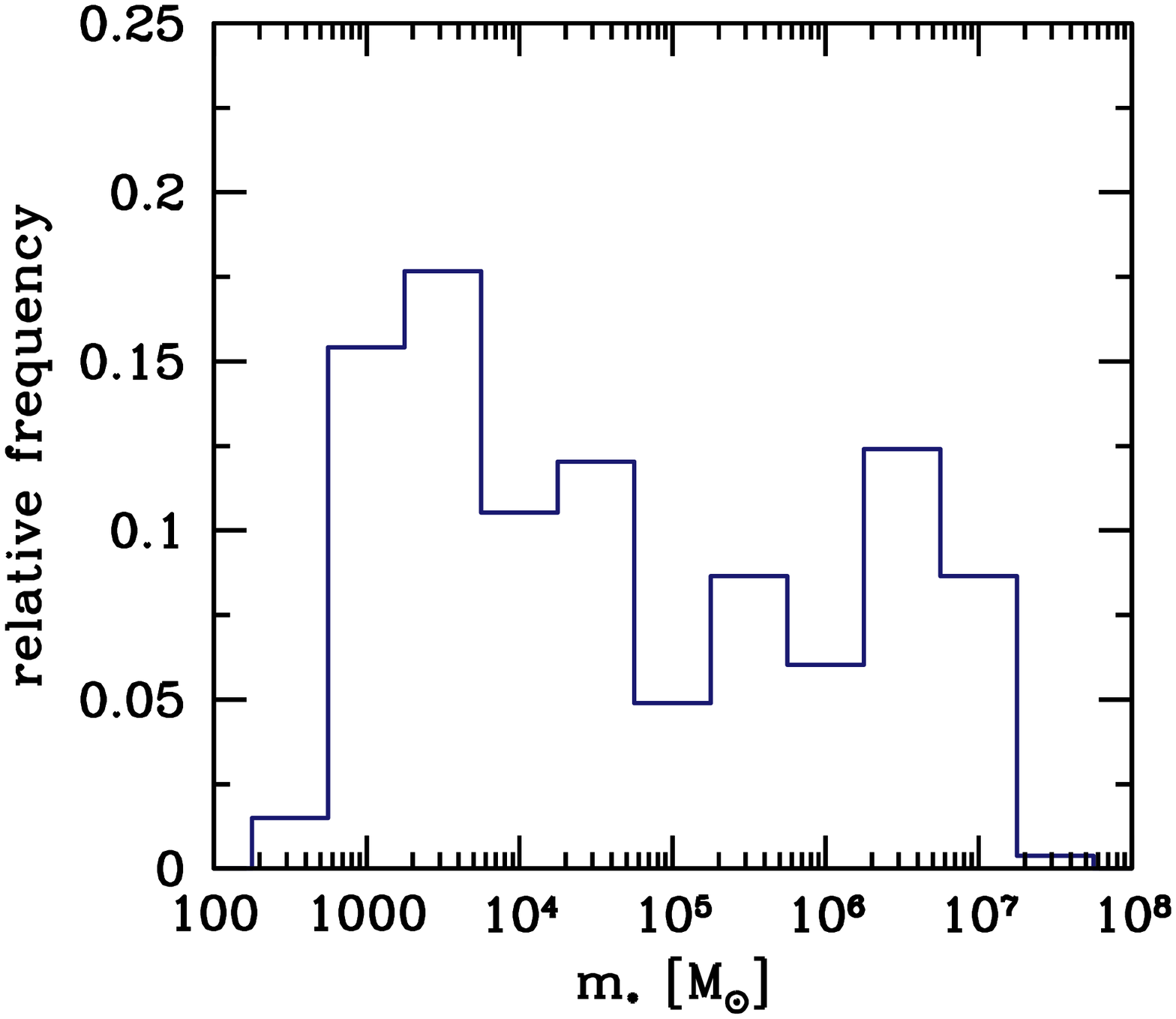}{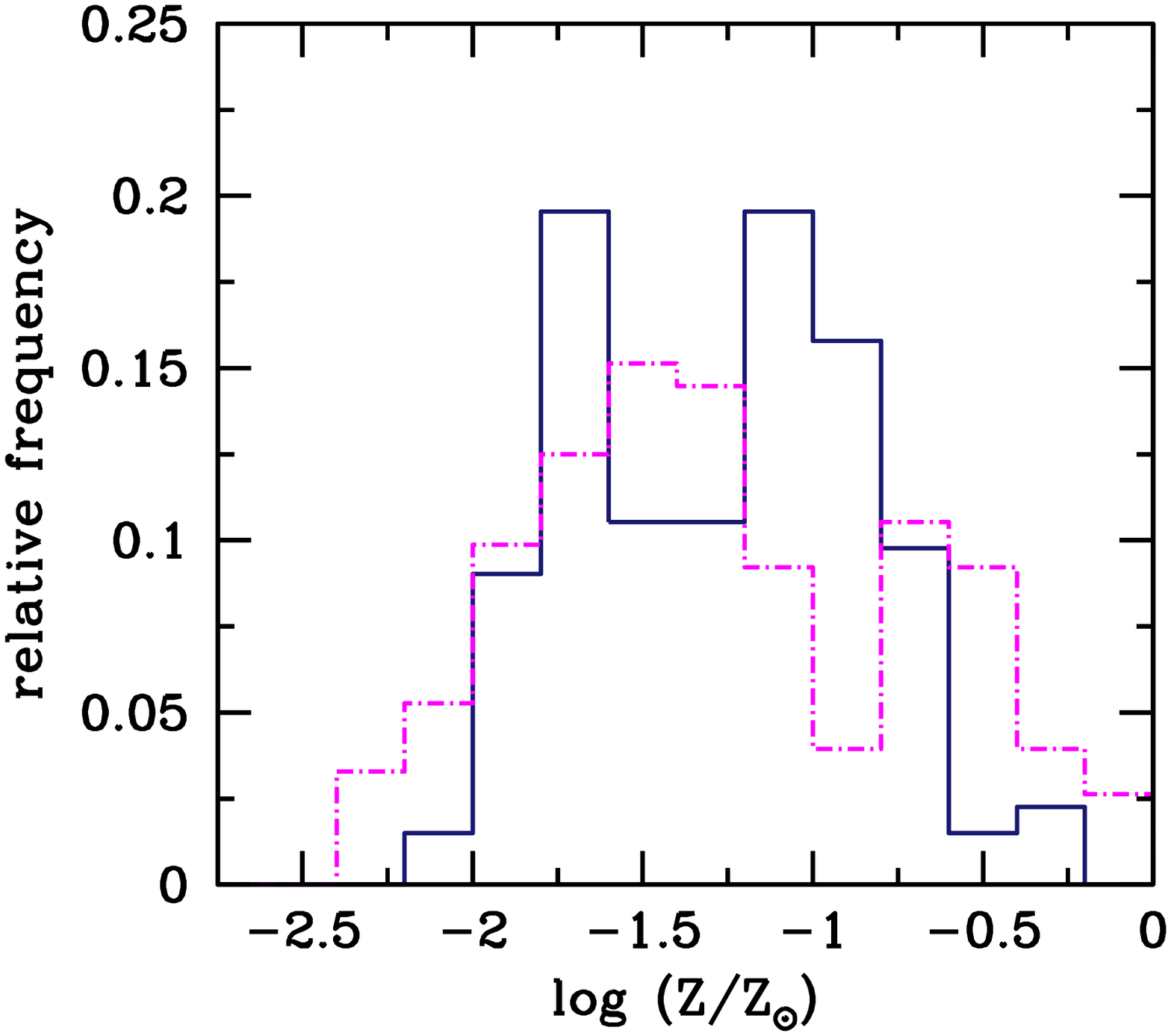}
\caption{Stellar mass (left panel) and gas-phase mean metallicity distribution (right panel) in our sample of colliding subhalo 
pairs (see text for details). The dashed histogram shows the observed distribution of stellar metallicities in 
Galactic globular clusters.
}
\label{figMZR}
\end{figure*}

\bigskip {\it (iv) The Metallicity of GCs:}
The GC population in the MW is observed to be clearly bimodal, with a low-metallicity component peaking at [Fe/H]$\simeq -1.55$, and a high-metallicity tail at $\simeq -0.55$ 
\citep{harris16}. Only 30\% of MW globulars have [Fe/H]$>-1$, but many massive galaxies possess strongly
bimodal GC systems, with nearly equal numbers of metal-rich and metal-poor clusters \citep{peng06}. Since stars within most GCs do not show an internal spread in iron-peak elements, there must 
exist a mechanism that chemically homogenizes the gas within a protocluster before the onset of star formation. The dominant mode of chemical mixing is thought to be turbulent diffusion \citep{murray90}, 
which has been shown to produce a stellar abundance scatter that is much smaller than that of the star-forming gas \citep{feng14}. 
In our model, GCs will inherit the gas-phase metallicity of the interacting subhalo pair that triggers their formation, and a useful perspective 
can be obtained by assigning a stellar mass to each subhalo at infall following the median stellar-to-halo mass relation from \citet{behroozi13}. This redshift-dependent prescription, extrapolated to the small scales of interest here, leads to a broad range of stellar masses, $10^{2.6}<m_*/\msun<10^{8.6}$ for our sample of colliding 
substructures, a distribution that extends over nearly 6 decades in mass with a median value equal to $10^{4.4}\,\msun$ (see Fig. \ref{figMZR}). 

A general tendency of decreasing metallicity towards lower stellar masses is commonly accepted, but the exact form of the stellar mass ($m_*$) vs. gas-phase 
metallicity ($Z$) relation (hereafter MZR) and its evolution with redshifts are currently poorly known as a result of the presence of strong systematic uncertainties 
affecting metallicity diagnostics. A few studies, mostly at $z\sim 0$, have tried to extend the MZR to the low-mass dwarf galaxy regime, deriving power-law relations
($Z\propto m_*^\alpha$) with slopes $\alpha=0.29\pm 0.03$ \citep{berg12}, $\simeq 0.5$ \citep{andrews13}, $0.44\pm 0.1$ \citep[][high star formation rate bin]{jimmy15}, 
and $0.14\pm 0.08$ \citep{blanc19}. Here, we adopt the intermediate-range value of $\alpha=0.37$ \citet{blanc19,ma15} with $Y$-intercept equal to 5.50, i.e.     
\begin{equation}
\begin{aligned}
\log (Z/Z_\odot) & = 12+\log ({\rm O/H})-8.69\\
& = 5.50+0.37\log (m_*/\msun)-8.69.
\end{aligned}
\end{equation}
The zero-point was chosen to give $12+\log ({\rm O/H})=8.18$ at $\log (m_*/\msun)=7.25$, in agreement with the DEEP2 low-mass MZR \citep{zahid12,blanc19}. 
Little is known about the redshift evolution of the MZR on dwarf-galaxy scales; in order to minimize the number of free parameters and for ease of interpretation, 
we assume no evolution with time in the following \citep[see also][]{hidalgo17}. 

The expected GC metallicity, computed by averaging the gas metallicities of each colliding pair, is shown in Figure 
\ref{figMZR}, together with the observed distribution of [Fe/H] in Galactic globulars, both metal-poor and metal-rich \citep{harris96}. Our simple scheme 
predicts a spread in metallicities that is similar to that observed, with a distribution that is strongly bimodal. In contrast to some previous work, we do 
not explicitly assume separate pathways for the formation of blue and red globulars, and simply predict the metallicity of each GC based on the enrichment level of the interacting 
subhalo pair that triggered its formation. About 30\% of the colliding pairs have metallity $\log (Z/Z_\odot)\ge -1$: the red globulars are the result of impacts that 
involve at least one massive, more chemically evolved satellite, and the metallicity bimodality reflects a bimodality in stellar and peak halo masses 
for the colliding subhalos.
The age spread is similar in both the blue and red populations, but the red and blue peaks are shifted towards lower values compared to the MW GC data. 

Given the intrinsic uncertainties in the stellar-to-halo mass relation, the MZR, and their evolution with redshift on small mass-galaxy scales,    
it seems again ill-advised to draw definite conclusions from this comparison. Evolutionary corrections on the MZR will shift the predicted values towards even lower metallicities.
We notice here two effects that have been ignored and may skew the predicted distribution in the opposite direction, towards higher metallicities: 1) collisions involving 
massive, enriched subhalos that, albeit rarer, could produce several GCs per event. The impact of this (unknown) multiplicity factor 
has been neglected in Figure \ref{figMZR}; and 2) high-speed bound collisions between subhalos falling in on radial 
orbits and the central most massive progenitor, which may result in a more centrally concentrated subpopulation of metal-rich 
globulars \citep[see also][]{griffen+10}. We count only 17 such collisions, compared to the 133 subhalo-subhalo impacts that satisfy 
our criteria. 

\bigskip {\it (v) The Spatial Distribution of GCs:} A common feature of many globular formation models is the reliance on some ad hoc assumptions to identify
the sites where GCs form. The median Galactocentric distance of all known MW (blue$+$red) GCs is 5 kpc \citep{harris96}, with the metal-poor, [Fe/H]$<-1$ subpopulation
being less spatially concentrated (median Galactocentric distance of 7.5 kpc). In our sample of colliding subhalos, we find that 80\% of all clumps
surviving complete disruption have first pericenter distances smaller than 20 kpc, with a median value of $12\,$ kpc (see the bottom right panel in Fig. \ref{figVLI}).
Models in which the bulk of the metal-poor GC subpopulation formed in satellite systems -- many of which are now tidally disrupted -- and were subsequently accreted onto the 
main galaxy tend to produce, however, clusters with a more extended spatial distribution than observed \citep{muratov10,creasey19}, 
unless they are associated with rare, early progenitor halos at $z\sim 10$ \citep{moore06,katz14}.
Our collision-driven scenario may offer a new mechanism for biasing the spatial distribution of GCs relative to the overall mass profile. This is because, in an
inelastic collision, the splash remnant will lose orbital energy and fall deeper into the Galactic potential rather than sharing the orbits of the progenitor
subhalos. It is interesting to briefly examine here the impact of kinetic energy dissipation on orbital parameters.
Consider, for simplicity, two subhalos moving on coplanar, coaxial, {\it prograde} elliptical orbits in a Keplerian potential of gravitational parameter $\mu$.
The two orbits have the same specific angular momentum $h$ and the subhalos collide at the semiparameter location of the ellipses, $p=h^2/\mu$.
Let us decompose the velocity vectors along the outward radial direction ($\hat r$) and the prograde azimuthal direction ($\hat \theta$) at this position.
The radial and azimuthal velocity components of the two subhalos just prior to impact are $(\mu e_1/h, \mu/h)$ and $(\mu e_2/h, \mu/h)$, respectively,
where $e_1$ and $e_2$ are the orbital eccentricities, with $e_1<e_2$. In the case of a perfectly inelastic encounter between two bodies of equal mass $m_c$, conservation of linear
momentum determines the postcollision instantaneous orbital velocity vector $\vec v_f$ at $p$ of the combined remnant as
\begin{equation}
\vec v_f={1\over 2}(\vec v_1+\vec v_2)={\mu\over 2h}~[(e_1+e_2)\hat r+2\hat \theta].
\end{equation}
In the absence of external torques the total angular momentum of the system remains unchanged, but the dissipation of kinetic energy leads to a net loss of
orbital energy
\begin{equation}
\begin{aligned}
\Delta E=-{m_c\mu^2\over 4h^2}(e_2-e_1)^2\le 0.\\
\\
\end{aligned}
\end{equation}
The new specific orbital energy is
\begin{equation}
\begin{aligned}
\epsilon={1\over 2} v_f^2 -{\mu^2\over h^2}={\mu^2\over 2h^2}\left[{(e_1+e_2)^2\over 4}-1\right],\\
\\
\end{aligned}
\end{equation}
and the new eccentricity is
\begin{equation}
e=\sqrt{1+2\epsilon_fh^2/\mu^2}={e_1+e_2\over 2}.
\end{equation}
Another simple situation to analyze is the case of two clumps moving on coplanar, coaxial prograde orbits with different angular momenta but same pericenter distance $r_p$.
Conservation of angular momentum in this case yields for the new eccentricity of the collision remnant
\begin{equation}
\sqrt{1+e}={\sqrt{1+e_1}+\sqrt{1+e_2}\over 2},
\end{equation}
and the change in orbital energy is again
\begin{equation}
\begin{aligned}
\Delta E=-{m_c\mu\over 2r_p}(e_2+e_1-2e)\le 0.\\
\\
\end{aligned}
\end{equation}
The above examples emphasize the fact that the remnant will initially be moving on a less energetic orbit with an eccentricity that is intermediate between those of the colliding pair.
One would expect globulars to progressively lose memory of their initial infall direction as they orbit in a host halo that is clumpy and triaxial. 
Detailed calculations of the expected radial profile and 3-D distribution of GCs are beyond the scope of this paper and are postponed to future work.

\acknowledgments \ni

We thank Z. Haiman and M. Krumholz for very helpful comments and suggestions on various aspects of this manuscript. 
Support for this work was provided by NASA through a contract to the WFIRST-EXPO Science Investigation Team (15-WFIRST15-0004), administered by the GSFC (P.M.).

%%%%%%%%%%%%%%%%%%%%%%%%%%%%%%%%%%%%%%%%%%%%%%%

\bibliographystyle{apj}
\bibliography{paper}

%%%%%%%%%%%%%%%%%%%%%%%%%%%%%%%%%%%%%%%%%%%%%%%

\label{lastpage}

\end{document}